\documentclass[12pt]{article}
\usepackage{mathrsfs}
\usepackage{graphicx}
\usepackage{amsmath}
\usepackage{amssymb}
\usepackage{caption2}
\begin{document}
\newcommand  {\ba} {\begin{eqnarray}}
\newcommand  {\be} {\begin{equation}}
\newcommand  {\ea} {\end{eqnarray}}
\newcommand  {\ee} {\end{equation}}
\renewcommand{\thefootnote}{\fnsymbol{footnote}}
\renewcommand{\figurename}{Figure.}
\renewcommand{\captionlabeldelim}{.~}

\vspace*{1cm}
\begin{center}
 {\Large\textbf{Baryogenesis and Asymmetric Dark Matter from The Left-Right Mirror Symmetric Model}}

\vspace{1cm}
 \textbf{Wei-Min Yang}

\vspace{0.4cm}
 \emph{Department of Modern Physics, University of Science and Technology of China, Hefei 230026, P. R. China}

\vspace{0.2cm}
 \emph{E-mail: wmyang@ustc.edu.cn}
\end{center}

\vspace{1cm}
\noindent\textbf{Abstract}: The paper suggests a left-right mirror symmetric model to account for the baryogenesis and
 asymmetric dark matter. The model can simultaneously accommodate the standard model, neutrino physics, matter-antimatter asymmetry and dark matter. In particular, it naturally and elegantly explains the origin of the baryon and dark matter asymmetries, and clearly gives the close interrelations of them. In addition, the model predicts a number of interesting results, e.g. the cold dark matter neutrino mass is $3.1$ times the proton mass. It is also feasible and promising to test the model in future experiments.

\vspace{1cm}

\vspace{0.3cm}

\newpage
 \noindent\textbf{I. Introduction}

\vspace{0.3cm}
 The standard model (SM) has been evidenced to be a very successful theory at the electroweak energy scale. The precise tests for the SM physics have established plenty of knowledge about the elementary particles \cite{1,2}. Nevertheless, at present there are a number of the unsolved issues in the particle physics and universe observations \cite{3,4}, which are not able to be accounted by the SM. The issues in flavor physics are the two facts, i) the mass spectrum hierarchy of the quarks and charged leptons \cite{5}, ii) the distinct difference between the quark flavor mixing pattern and the lepton one \cite{6}. The neutrino physics has to answer such questions as the real origin of the Sub-eV neutrino masses \cite{7}? Whether the nature of the light neutrinos are Dirac or Majorana fermions \cite{8}? Is the $CP$ violation in the lepton mixing vanishing or not \cite{9}? However, the issues in the cosmology are more difficult. What is real mechanism of the genesis of the matter-antimatter asymmetry \cite{10}? What is the nature of the cold dark matter \cite{11}? The current universe observations have given the data of the baryon asymmetry and the relic abundance of the cold dark matter as follows \cite{1},
\begin{alignat}{1}
 \eta_{B}=\frac{n_{B}-\overline{n}_{B}}{n_{\gamma}}\approx6.15\times10^{-10},\hspace{0.5cm}\frac{\Omega_{D}}{\Omega_{B}}\approx5.
\end{alignat}
 In particular, the two issues about the baryogenesis and dark matter are very significant for both particle physics and cosmology, so they attract great attentions in the experiment and theory fields all the time \cite{12}.

 The various theoretical suggestions have been proposed to solve the above-mentioned problems \cite{13}. The baryogenesis can be achieved by the electroweak baryogenesis \cite{14}, the leptogenesis \cite{15}, the Dirac leptogenesis \cite{16}, and so on. The cold dark matter candidates are possible the scalar boson dark matter \cite{17}, the sterile neutrino dark matter \cite{18}, the supersymmetry dark matter\cite{19}, and so on \cite{20}. The asymmetric dark matter ideas have been discussed in the references \cite{21}. The references \cite{22} has studied the mirror symmetric model. These theories are successful in explaining one specific aspect of the problems, but it seems very difficult for them to solve many aspects of the problems simultaneously. On the basis of the unity of nature, a realistic theory beyond the SM should simultaneously accommodate and account for the neutrino physics, baryon asymmetry and dark matter besides the SM, in other words, it has to integrate the four things completely. It is especially hard for a model construction to keep the principle of the simplicity, feasibility and the fewer number of parameters, otherwise, the theory will be excessive complexity and incredible or infeasibility. However, it is still a large challenge for theoretical particle physicists to realize the purpose \cite{23}.

 In this work, I construct a simple and feasible particle model. It can simultaneously accommodate the SM, neutrino physics, matter-antimatter asymmetry and dark matter. The model extends the SM to a left-right mirror symmetry theory. It has  the local gauge groups $SU(3)_{C}\otimes SU(2)_{L}\otimes SU(2)_{R}\otimes U(1)_{Y}$ and global symmetry $U(1)_{B+\widetilde{B}}\otimes U(1)_{L}\otimes U(1)_{\widetilde{L}}$ where $\widetilde{B}$ and $\widetilde{L}$ are respectively the mirror baryon and lepton numbers. In addition, the model has the left-right mirror symmetry and a $Z_{2}$ discrete symmetry. Besides the SM particles, the model introduces the corresponding mirror particles. The model symmetries are spontaneously broken step by step at different energy scale as the universe temperature decreasing. In the model, the super-heavy scalar boson can decay into the SM right-handed quarks and the left-handed mirror quarks. The decay processes are out-of-equilibrium and $CP$ violation. The $CP$-violating source lies in the explicit mirror breaking coupling in the Yukawa sector. This eventually leads to both the baryon asymmetry and the mirror neutrino asymmetry through the two steps of the sphaleron processes \cite{24}. The lightest mirror neutrino, which is a Dirac neutrino with the GeV mass, is exactly the cold dark matter. The model can not only completely accommodate the SM and neutrino physics, but also correctly reproduce the observed value of the baryon asymmetry and the relic abundance of the cold dark matter. In particular, the model predicts some interesting results, for example, the cold dark matter asymmetry is $1.6$ times the baryon asymmetry, its mass is $3.1$ times the proton mass, and so on. Finally, the model is feasible and promising to be tested in future experiments. I give some methods of searching some of the new particles at the colliders.

 The remainder of this paper is organized as follows. In Section II I outline the model. Sec. III I discuss the matter-antimatter asymmetry and dark matter. The numerical results and the experimental searches are given by Sec. IV. Sec. V is devoted to conclusions.

\vspace{1cm}
 \noindent\textbf{II. Model}

\vspace{0.3cm}
 The gauge symmetries of the model are characterized by the Local and global gauge groups as follows,
\begin{alignat}{1}
 &\mbox{Local gauge groups}: SU(3)_{C}\otimes SU(2)_{L}\otimes SU(2)_{R}\otimes U(1)_{Y},\nonumber\\
 &\mbox{Global gauge groups}: U(1)_{B+\widetilde{B}}\otimes U(1)_{L}\otimes U(1)_{\widetilde{L}},
\end{alignat}
 where $\widetilde{B}$ and $\widetilde{L}$ respectively indicate the baryon and lepton number of the mirror particles. The gauge groups have evidently a left-right mirror symmetry. The model particle contents and their gauge quantum numbers are in detail listed as follows,
\begin{alignat}{1}
 &G_{\mu}^{a}(8,1,1,0),\hspace{0.5cm}W_{L\mu}^{i}(1,3,1,0),\hspace{0.5cm}W_{R\mu}^{i}(1,1,3,0),\hspace{0.5cm}B_{\mu}(1,1,1,1),\nonumber\\
 &q_{L}(3,2,1,\frac{1}{3}),\hspace{0.5cm} \widetilde{q}_{R}(3,1,2,\frac{1}{3}),\hspace{0.5cm}
  (u_{R},\widetilde{u}_{L})(3,1,1,\frac{4}{3}),\hspace{0.5cm} (d_{R},\widetilde{d}_{L})(3,1,1,-\frac{2}{3}),\nonumber\\
 &l_{L}(1,2,1,-1),\hspace{0.5cm} \widetilde{l}_{R}(1,1,2,-1),\hspace{0.5cm}
  (\nu_{R},\widetilde{\nu}_{L})(1,1,1,0),\hspace{0.5cm} (e_{R},\widetilde{e}_{L})(1,1,1,-2),\nonumber\\
 &H_{L}(1,2,1,1),\hspace{0.5cm} H_{R}(1,1,2,1),\hspace{0.5cm}
  S(1,1,1,0)_{L=-2},\hspace{0.5cm} \widetilde{S}(1,1,1,0)_{\widetilde{L}=-2},\nonumber\\
 &\phi^{\pm}(1,1,1,\pm2),\hspace{0.5cm} \phi^{0}(1,1,1,0).
\end{alignat}
 These notations are self-explanatory. All kinds of the fermions imply three generations as usual. $\nu_{R}$ and $\widetilde{\nu}_{L}$ are singlet neutrinos. $S$ and $\widetilde{S}$ are complex singlet bosons with the lepton number $(-2)$. $\phi^{\pm}$ and $\phi^{0}$ are super-heavy bosons, moreover, $\phi^{0}$ is a real scalar.

 In addition to the above-mentioned gauge symmetries, the model has a attractive left-right mirror symmetry and a $Z_{2}$ discrete symmetry. They are defined by the field transforms as follows,
\begin{alignat}{1}
 \mbox{SM sector}:\hspace{0.2cm}
 &q_{L},\hspace{0.2cm}u_{R},\hspace{0.2cm}d_{R},\hspace{0.2cm}l_{L},\hspace{0.2cm}\nu_{R},\hspace{0.2cm}e_{R},\hspace{0.2cm}
  H_{L},\hspace{0.2cm}S,\hspace{0.2cm}\phi^{\pm},\hspace{0.2cm}\phi^{0},\hspace{0.2cm}
  W_{L\mu},\hspace{0.2cm}G_{\mu},\hspace{0.2cm}B_{\mu}.\nonumber\\
 &\updownarrow\hspace{4.7cm}\updownarrow\hspace{3.1cm}\updownarrow \nonumber\\
  \mbox{Mirror sector}:\hspace{0.2cm}
 &\widetilde{q}_{R},\hspace{0.2cm}\widetilde{u}_{L},\hspace{0.2cm}\widetilde{d}_{L},\hspace{0.2cm}
  \widetilde{l}_{R},\hspace{0.2cm}\widetilde{\nu}_{L},\hspace{0.2cm}\widetilde{e}_{L},\hspace{0.2cm}
  H_{R},\hspace{0.2cm}\widetilde{S},\hspace{0.2cm}\phi^{\pm},\hspace{0.2cm}\phi^{0},\hspace{0.2cm}
  W_{R\mu},\hspace{0.2cm}G_{\mu},\hspace{0.2cm}B_{\mu}.\nonumber\\
  \mbox{$Z_{2}$ parity is``$+1$"}&\:\mbox{for}\:f,\:H_{L},\:H_{R},\:S,\:\widetilde{S}\:
   \mbox{and all the gaugebosons}.\nonumber\\
  \mbox{$Z_{2}$ parity is``$-1$"}&\:\mbox{for}\:\widetilde{f},\:\phi^{\pm},\:\phi^{0}.
\end{alignat}
 Here the called `` SM sector" not only contains all the SM particles but also it introduce the non-SM particles as $\nu_{R},S$. For the four particles, $\phi^{\pm},\phi^{0},G_{\mu},B_{\mu}$, their mirror particles are exactly themselves, so they are actually shared in the two sectors. The gauge and discrete symmetries will be broken step by step at different energy scales in the evolution of the universe.

 On the basic of the above symmetries, the model Lagrangian is composed of the following three parts. Firstly, the gauge kinetic energy terms are
\begin{alignat}{1}
 \mathscr{L}_{Gauge}=
 &\:\mathscr{L}_{pure\,gauge}+\sum_{f}i\overline{f}\gamma^{\mu}D_{\mu}f
  +\sum_{\widetilde{f}}i\overline{\widetilde{f}}\gamma^{\mu}D_{\mu}\widetilde{f}\nonumber\\
 &+(D^{\mu}H_{L})^{\dagger}(D_{\mu}H_{L})+(D^{\mu}H_{R})^{\dagger}(D_{\mu}H_{R})
  +(\partial^{\mu}S)^{\dagger}\partial_{\mu}S+(\partial^{\mu}\widetilde{S})^{\dagger}\partial_{\mu}\widetilde{S}\nonumber\\
 &+(D^{\mu}\phi^{\pm})^{\dagger}(D_{\mu}\phi^{\pm})+\frac{1}{2}\partial^{\mu}\phi^{0}\partial_{\mu}\phi^{0},
\end{alignat}
 where $f$ and $\widetilde{f}$ denote the SM and mirror fermions in (4), respectively. The covariant derivative $D_{\mu}$ is given by
\ba
 D_{\mu}=\partial_{\mu}+i\left(g_{s}G_{\mu}^{a}\frac{\lambda^{a}}{2}+g_{w}W_{L\mu}^{i}\frac{\tau_{L}^{i}}{2}
 +g_{w}W_{R\mu}^{i}\frac{\tau_{R}^{i}}{2}+g_{Y}B_{\mu}\frac{Q_{Y}}{2}\right),
\ea
 where $g_{s},g_{w},g_{Y}$ are three gauge coupling constants, $\lambda^{a}$ and $\tau^{i}$ are respectively Gell-Mann and Pauli matrices, and $Q_{Y}$ is the charge operator of $U(1)_{Y}$.

 Secondly, the model Yukawa couplings are
\begin{alignat}{1}
  \mathscr{L}_{Yukawa}=
 &\hspace{0.3cm}\overline{q_{L}}H_{L}'Y_{u}u_{R}+\overline{q_{L}}H_{L}Y_{d}d_{R}
   +\overline{l_{L}}H_{L}Y_{e}e_{R}+\overline{l_{L}}H_{L}'Y_{\nu}\nu_{R}+S\nu_{R}^{\,T}CY_{m}\nu_{R}\nonumber\\
 &+\overline{\widetilde{q}_{R}}H_{R}'Y_{u}\widetilde{u}_{L}+\overline{\widetilde{q}_{R}}H_{R}Y_{d}\widetilde{d}_{L}
  +\overline{\widetilde{l}_{R}}H_{R}Y_{e}\widetilde{e}_{L}+\overline{\widetilde{l}_{R}}H_{R}'Y_{\nu}\widetilde{\nu}_{L}
  +\widetilde{S}\widetilde{\nu}_{L}^{\,T}CY_{m}\widetilde{\nu}_{L}\nonumber\\
 &+\phi^{+}\overline{u_{R}}Y_{0}\widetilde{d}_{L}+\phi^{+}\overline{\widetilde{u}_{L}}Y_{0}^{*}d_{R}
  +\phi^{0}\overline{u_{R}}Y_{1}\widetilde{u}_{L}+\phi^{0}\overline{d_{R}}Y_{2}\widetilde{d}_{L}+h.c.\,,
\end{alignat}
 where $H_{L/R}'=i\tau_{2}H_{L/R}^{*}$ and $C$ is a charge conjugation matrix. Obviously, the parameter freedom is greatly reduced owing to the mirror symmetry. The couplings, $Y_{u},Y_{d}, etc.$, are $3\times3$ complex matrices but $Y_{1},Y_{2}$ are Hermitian matrices. In any case, I can always choose such a set of flavor basic that $Y_{e},Y_{m},Y_{1},Y_{2}$ are all diagonal matrices and the others are all non-diagonal. The coupling matrices should originate from family symmetry breaking, however, they bring about flavor mixings. In addition, the irremovable complex phases in the couplings become sources of the $CP$ violation. In particular, the irremovable complex phase in $Y_{0}$ is also a source of the mirror symmetry breaking, explicitly, the mirror symmetry is broken for $Y_{0}\neq Y_{0}^{*}$. The Yukawa couplings of (7) will lead to reasonable explanations for the neutrino masses, matter-antimatter asymmetry and cold dark matter.

 Thirdly, the model scalar potential is given by
\begin{alignat}{1}
 V_{Scalar}=
 &\:\lambda_{H}\left(H_{L}^{\dagger}H_{L}-\frac{v_{L}^{2}}{2}+\frac{\mu_{0}v_{\phi}}{2\lambda_{H}}\right)^{2}
  +\lambda_{H}\left(H_{R}^{\dagger}H_{R}-\frac{v_{R}^{2}}{2}+\frac{\mu_{0}v_{\phi}}{2\lambda_{H}}\right)^{2}\nonumber\\
 &+\lambda_{S}\left(S^{\dagger}S-\frac{v_{s}^{2}}{2}\right)^{2}+\lambda_{S}\left(\widetilde{S}^{\dagger}\widetilde{S}\right)^{2}\nonumber\\
 &+\frac{\lambda_{\phi^{0}}}{4}\left(\phi^{02}-v_{\phi}^{2}+\frac{M_{\phi^{0}}^{2}}{\lambda_{\phi^{0}}}\right)^{2}
  +\lambda_{\phi^{\pm}}\left(\phi^{+}\phi^{-}+\frac{M_{\phi^{\pm}}^{2}}{2\lambda_{\phi^{\pm}}}\right)^{2}\nonumber\\
 &-\mu_{0}\phi^{0}\left(H_{L}^{\dagger}H_{L}+H_{R}^{\dagger}H_{R}\right)+\mbox{other weak coupling terms}.
\end{alignat}
 The self-coupling parameters, $\lambda_{H},\lambda_{S},\lambda_{\phi^{0}},\lambda_{\phi^{\pm}}$, are positive and should be $\sim0.1$. $\mu_{0}$ is a positive coupling parameter with mass dimension, explicitly, the $Z_{2}$ discrete symmetry is  broken by the $\mu_{0}$ term. The other interactive couplings should be sufficient weak. $v_{L},v_{R},v_{s},v_{\phi}$ are respectively the VEVs of the corresponding scalar fields, but $\widetilde{S}$ and $\phi^{\pm}$ have vanishing VEVs, see the following equation (9). $M_{\phi^{0}}$ and $M_{\phi^{\pm}}$ are respectively the super-heavy masses of $\phi^{0}$ and $\phi^{\pm}$, which are $\sim10^{10}$ GeV. It can clearly be seen from (8) that $v_{L}\neq v_{R}\neq 0$ will spontaneously break the local gauge symmetries, while $v_{s}\neq 0$ will spontaneously break the global gauge symmetry $U(1)_{L}$, but $U(1)_{\widetilde{L}}$ is unbroken due to $v_{\widetilde{s}}=0$. Obviously, the gauge symmetry breakings simultaneously trigger the spontaneous breaking of the mirror symmetry. After the local gauge symmetry breaking, $\phi^{0}$ will also be induced to develop a small $v_{\phi}$ by the $\mu_{0}$ term. In short, the model scalar sector undertakes and implements the gauge and discrete symmetry breakings. It is more varied and interesting in comparison with the SM or MSSM Higgs sector \cite{25}.

 The potential vacuum configurations and the scalar boson masses are derived by the standard program. The detailed results are as follows,
\begin{alignat}{1}
 &\langle H_{L}\rangle=\frac{v_{L}}{\sqrt{2}}\,,\hspace{0.5cm}\langle H_{R}\rangle=\frac{v_{R}}{\sqrt{2}}\,,\hspace{0.5cm}
  \langle S\rangle=\frac{v_{s}}{\sqrt{2}}\,,\nonumber\\
 &\langle\phi^{0}\rangle=v_{\phi}=\frac{\mu_{0}(v_{L}^{2}+v_{R}^{2})}{2M_{\phi^{0}}^{2}}\,,\hspace{0.5cm}
  \langle \widetilde{S}\rangle=\langle \phi^{\pm}\rangle=0\,,\nonumber\\
 &M_{H_{L}}^{2}=2\lambda_{H}v_{L}^{2}+\mu_{0}v_{\phi},\hspace{0.5cm}M_{H_{R}}^{2}=2\lambda_{H}v_{R}^{2}+\mu_{0}v_{\phi},\nonumber\\
 &M_{S_{R}}^{2}=2\lambda_{S}v_{s}^{2},\hspace{0.5cm} M_{S_{I}}=M_{\widetilde{S}}=0,
\end{alignat}
 where $S_{R}$ and $S_{I}$ are respectively real and imaginary components of $S$. There are three massive neutral bosons $H_{R}^{0},H_{L}^{0},S_{R}$, while $S_{I}$ and $\widetilde{S}$ become massless Goldstone bosons. Some parameters in (9) are estimated as $v_{R}\sim10^{8},v_{L}\sim246,v_{s}\sim 100,\mu_{0}\sim 10^{3}$ (all are in GeV as unit), $v_{\phi}\sim0.1$ GeV is not a independent parameter since $M_{\phi^{0}}$ is regarded to be independent. This hierarchy of the VEVs clearly shows the symmetry breaking sequence. First of all, $SU(2)_{R}\otimes U(1)_{Y}$ is spontaneously broken down $U(1)_{Y'}$ which is namely the hypercharge subgroup of the SM. This is achieved by the neutral component $H_{R}^{0}$ developing $v_{R}$. Secondly, $SU(2)_{L}\otimes U(1)_{Y'}\rightarrow U(1)_{em}$, i.e. the electroweak breaking. This is accomplished by the neutral component $H_{L}^{0}$ developing $v_{L}$. Lastly, the real component $S_{R}$ developing $v_{s}$ completes the $U(1)_{L}$ breaking, whereas the $U(1)_{\widetilde{L}}$ symmetry is maintained and unbroken all the time. At the present day, $M_{H_{L}}$ has been measured as $125$ GeV at the LHC \cite{26}. The model predicts that $M_{H_{R}}$ is $\sim10^{8}$ GeV, $M_{S_{R}}$ is scores of GeVs, and the Goldstone bosons $S_{I}$ and $\widetilde{S}$ are actually a species of hot dark matter. However, the other bosons remain to be searched in future experiments.

 In the gauge sector, the local gauge symmetry breakings result in masses and mixings of the gauge fields through the Higgs mechanism. The detailed expressions are as follows,
\begin{alignat}{1}
 &g_{w}(W_{L\mu}^{i}\frac{\tau_{L}^{i}}{2}+W_{R\mu}^{i}\frac{\tau_{R}^{i}}{2})+g_{Y}B_{\mu}\frac{Q_{Y}}{2}\longrightarrow\nonumber\\
 &\frac{g_{w}}{\sqrt{2}}(W_{L\mu}^{+}\tau_{L}^{+}+W_{L\mu}^{-}\tau_{L}^{-}+W_{R\mu}^{+}\tau_{R}^{+}+W_{R\mu}^{-}\tau_{R}^{-})
  +g_{w}(Z_{\mu}Q_{w}+\widetilde{Z}_{\mu}\widetilde{Q}_{w})+eA_{\mu}Q_{e},\nonumber\\
 &tan\widetilde{\theta}=\frac{g_{Y}}{g_{w}}\,,\hspace{0.5cm}
  tan\theta_{W}=sin\widetilde{\theta}\,,\hspace{0.5cm} e=g_{w}sin\theta_{W},\nonumber\\
 &Q_{e}=I^{L}_{3}+I^{R}_{3}+\frac{Q_{Y}}{2}\,,\hspace{0.3cm}
  Q_{w}=\frac{I^{L}_{3}-sin^{2}{\theta_{W}}Q_{e}}{cos\theta_{W}}\,,\hspace{0.3cm}
  \widetilde{Q}_{w}=\frac{I^{R}_{3}-sin^{2}{\widetilde{\theta}}(Q_{e}-I^{L}_{3})}{cos\widetilde{\theta}}\,,\nonumber\\
 &\left(\begin{array}{c}A_{\mu}\\Z_{\mu}\\\widetilde{Z}_{\mu}\end{array}\right)
  =\left(\begin{array}{ccc}cos\theta&sin\theta&0\\-sin\theta&cos\theta&0\\0&0&1\end{array}\right)
  \left(\begin{array}{ccc}cos\widetilde{\theta}&0&sin\widetilde{\theta}\\0&1&0\\-sin\widetilde{\theta}&0&cos\widetilde{\theta}\end{array}\right)
  \left(\begin{array}{c}B_{\mu}\\W^{3}_{L\mu}\\W^{3}_{R\mu}\end{array}\right),\nonumber\\
 &M_{W_{L}}=\frac{g_{w}v_{L}}{2}\,,\hspace{0.3cm}M_{W_{R}}=\frac{g_{w}v_{R}}{2}\,,\hspace{0.3cm}
  M_{Z}=\frac{M_{W_{L}}}{cos\theta_{W}}\,,\hspace{0.3cm}
  M_{\widetilde{Z}}=\frac{M_{W_{R}}}{cos\widetilde{\theta}}\,,\hspace{0.3cm}
  M_{A_{\mu}}=0\,.
\end{alignat}
 In (10), there are only two independent parameters, i.e. $g_{w}$ and $g_{Y}$. $\widetilde{\theta}$ is a mixing angle for the $SU(2)_{R}$ breaking, while $\theta_{W}$ is a mixing angle for the $SU(2)_{L}$ breaking. The two angles are correlated by that equation in (10). The SM hypercharge is $\frac{Q_{Y}'}{2}=I^{R}_{3}+\frac{Q_{Y}}{2}$. $Q_{w}$ and $\widetilde{Q}_{w}$ are two charge operators associated with the two massive neutral gauge fields $Z_{\mu}$ and $\widetilde{Z}_{\mu}$, respectively. It should be pointed out that the mixing angle between $Z_{\mu}$ and $\widetilde{Z}_{\mu}$ is $\sim\frac{v_{L}^{2}cos\widetilde{\theta}\,tan^{2}\theta_{W}}{v_{R}^{2}cos\theta_{W}}$, it is so small that it can be ignored. For $v_{R}\sim 10^{8}$  GeV, $M_{W_{R}}$ and $M_{\widetilde{Z}}$ are $\sim 10^{7}$ GeV. They are too heavy to be detected at the present.

 In the Yukawa sector, the fermion masses and mixings are given as follows,
\begin{alignat}{1}
 &\mathscr{L}_{Mass}=
  (\overline{u_{L}},\overline{\widetilde{u}_{L}})
  \left(\begin{array}{cc}\frac{v_{L}}{\sqrt{2}}Y_{u}&0\\v_{\phi}Y_{1}^{\dagger}&\frac{v_{R}}{\sqrt{2}}Y_{u}^{\dagger}\end{array}\right)
  \left(\begin{array}{c}u_{R}\\\widetilde{u}_{R}\end{array}\right)
  +(\overline{d_{L}},\overline{\widetilde{d}_{L}})
  \left(\begin{array}{cc}\frac{v_{L}}{\sqrt{2}}Y_{d}&0\\v_{\phi}Y_{2}^{\dagger}&\frac{v_{R}}{\sqrt{2}}Y_{d}^{\dagger}\end{array}\right)
  \left(\begin{array}{c}d_{R}\\\widetilde{d}_{R}\end{array}\right)\nonumber\\
 &\hspace{1.5cm}+(\overline{e_{L}},\overline{\widetilde{e}_{L}})
  \left(\begin{array}{cc}\frac{v_{L}}{\sqrt{2}}Y_{e}&0\\0&\frac{v_{R}}{\sqrt{2}}Y_{e}^{\dagger}\end{array}\right)
  \left(\begin{array}{c}e_{R}\\\widetilde{e}_{R}\end{array}\right)
  +(\overline{\nu_{L}},\overline{\widetilde{\nu}_{L}})
  \left(\begin{array}{cc}\frac{v_{L}}{\sqrt{2}}Y_{\nu}&0\\0&\frac{v_{R}}{\sqrt{2}}Y_{\nu}^{\dagger}\end{array}\right)
  \left(\begin{array}{c}\nu_{R}\\\widetilde{\nu}_{R}\end{array}\right)\nonumber\\
 &\hspace{1.5cm}+\frac{1}{2}\nu_{R}^{\,T}C(\sqrt{2}v_{s}Y_{m})\nu_{R},\nonumber\\
 &\hspace{0.5cm}\mbox{the mixing angle of $u_{R}$ and $\widetilde{u}_{R}$ ($d_{R}$ and $\widetilde{d}_{R}$) is}
  \sim\frac{v_{\phi}Y_{1}}{v_{R}Y_{u}}\:(\frac{v_{\phi}Y_{2}}{v_{R}Y_{d}})\ll1,\nonumber\\
 &\hspace{0.5cm}M_{f=u,d,e,\nu}=-\frac{v_{L}}{\sqrt{2}}Y_{f}
  =U_{f_{L}}\mathrm{diag}\left(m_{f_{1}},m_{f_{2}},m_{f_{3}}\right)U^{\dagger}_{f_{R}},\nonumber\\
 &\hspace{0.5cm}M_{\widetilde{f}=\widetilde{u},\widetilde{d},\widetilde{e},\widetilde{\nu}}
  =-\frac{v_{R}}{\sqrt{2}}Y_{f}^{\dagger}=\frac{v_{R}}{v_{L}}M_{f}^{\dagger},\nonumber\\
 &\hspace{0.5cm}M_{\nu_{R}}=-\sqrt{2}\,v_{s}Y_{m},\hspace{0.3cm}M_{\nu_{L}}^{eff}=-M_{\nu}M_{\nu_{R}}^{-1}M_{\nu}^{T}
  =U_{\nu_{L}}^{eff}\mathrm{diag}\left(m_{\nu_{L1}},m_{\nu_{L2}},m_{\nu_{L3}}\right)U_{\nu_{L}}^{eff\,T},\nonumber\\
 &\hspace{0.5cm}U_{CKM}=U^{\dagger}_{u_{L}}U_{d_{L}},\hspace{0.5cm}U_{PMNS}=U^{\dagger}_{e_{L}}U_{\nu_{L}}^{eff},\nonumber\\
 &\hspace{0.5cm}\widetilde{U}_{CKM}=U^{\dagger}_{\widetilde{u}_{R}}U_{\widetilde{d}_{R}}=U_{CKM},\hspace{0.5cm}
  \widetilde{U}_{PMNS}=U^{\dagger}_{\widetilde{e}_{R}}U_{\widetilde{\nu}_{R}}\neq U_{PMNS}.
\end{alignat}
 By virtue of the small mixing of $u_{R}$ and $\widetilde{u}_{R}$ ($d_{R}$ and $\widetilde{d}_{R}$), the mirror quarks can oscillate into the SM quarks. This plays key roles in the following baryogenesis. As a result of the $U(1)_{L}$ breaking, $\nu_{R}$ obtains a Majorana mass about a dozen GeV, it becomes a heavy Majorana neutrino undetected by now. For $Y_{\nu}\sim10^{-7}$, $\nu_{L}$ is generated an effective Majorana mass through the see-saw mechanism \cite{27}, it is exactly Sub-eV Majorana neutrino in nature. By contrast, the mirror neutrino $\widetilde{\nu}$ has only a Dirac mass due to $U(1)_{\widetilde{L}}$ being unbroken, so it is actually a Dirac neutrino. The $\widetilde{\nu}$ masses are several to dozens of GeVs, in particular, the lightest mirror neutrino $\widetilde{\nu}_{1}$ becomes the cold dark matter. This distinction between the $L$ sector and the $\widetilde{L}$ one also leads to $\widetilde{U}_{PMNS}\neq U_{PMNS}$. The flavor mixing matrices $U_{CKM}$ and $U_{PMNS}$ are respectively defined by \cite{28,29}. The mixing angles and $CP$-violating phases in the two unitary matrices are parameterized by the standard form in particle data group \cite{1}.

 In conclusion, the above contents form the theoretical framework of the model. In brief, the model extends the SM to the left-right mirror symmetrical theory. The new symmetries and non-SM particles will play key roles in the new physics beyond the SM, in particular, in the origin of matter-antimatter asymmetry and cold dark matter.

\vspace{1cm}
 \noindent\textbf{III. Baryogenesis and Asymmetric Dark Matter}

\vspace{0.3cm}
 The baryogenesis and asymmetric dark matter have a common origin in the model, so the two things have a close relationship. As the universe expansion and cooling, the model symmetries are spontaneously broken and reduced step by step. In the evolution process, the baryogenesis and asymmetric dark matter will naturally be generated by the following mechanism.

 After the universe inflation, the universe reheating temperature is in general $\sim 10^{12-13}$ GeV for most of the inflation models \cite{30}. In the reheated universe, thus there is an immense amount of the super-heavy scalar boson $\phi^{0}$ which has by nature a mass about $10^{10}$ GeV. In the light of (7) and (8), The decay channels of $\phi^{0}$ include $\phi^{0}\rightarrow(u_{R}+\overline{\widetilde{u}_{L}})/(d_{R}+\overline{\widetilde{d}_{L}})$ and $\phi^{0}\rightarrow(H_{L}+H_{L}^{\dagger})/(H_{R}+H_{R}^{\dagger})$, but the main decay are actually the former modes and their $CP$ conjugate processes, as shown in Figure 1.
\begin{figure}
 \centering
 \includegraphics[totalheight=4.5cm]{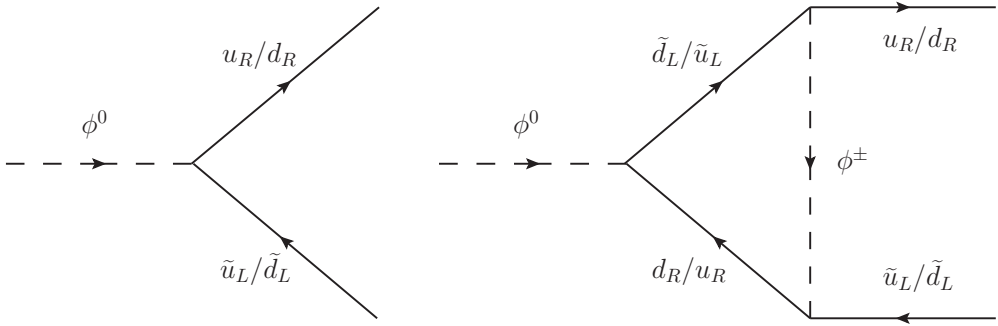}
 \caption{The tree and loop diagrams of the decays $\phi^{0}\rightarrow(u_{R}+\overline{\widetilde{u}_{L}})/(d_{R}+\overline{\widetilde{d}_{L}})$, which lead to the matter-antimatter asymmetry.}
\end{figure}
 For the couplings $Y_{1,2}\sim10^{-4}$, the decay rates are far smaller than the Hubble expansion rate of the universe, namely
\begin{alignat}{1}
 \Gamma(\phi^{0}\rightarrow u_{R}+\overline{\widetilde{u}_{L}}/d_{R}+\overline{\widetilde{d}_{L}})
 &=\frac{M_{\phi^{0}}Tr(Y_{i}Y_{i}^{\dagger})}{16\pi}\nonumber\\
 &\ll H(T=M_{\phi^{0}})=\frac{1.66\sqrt{g_{*}}M_{\phi^{0}}^{2}}{M_{pl}}\,,
\end{alignat}
 where $i=1,2$, $M_{pl}=1.22\times10^{19}$ GeV, $g_{*}$ is an effective number of relativistic degrees of freedom at $T=M_{\phi^{0}}$. At this temperature, the non-relativistic particles are only $\phi^{0}$ and $\phi^{\pm}$ in the model, the rest of the model particles are all relativistic states, so one can easy figure out $g_{*}=210$. Consequently, the $\phi^{0}$ decays are actually out-of-equilibrium processes.

 The complex phase in the coupling $Y_{0}$ is explicitly a source of the $CP$ violation and mirror symmetry breaking. It can surely lead to $CP$ asymmetries of the decays by the interference between the tree diagram and the loop one. The $CP$ asymmetries are defined and calculated as follow,
\begin{alignat}{1}
 &\Gamma_{total}(\phi^{0})=\Gamma(\phi^{0}\rightarrow u_{R}+\overline{\widetilde{u}_{L}})
   +\Gamma(\phi^{0}\rightarrow\overline{u_{R}}+\widetilde{u}_{L})\nonumber\\
 &\hspace{2cm}+\Gamma(\phi^{0}\rightarrow d_{R}+\overline{\widetilde{d}_{L}})
  +\Gamma(\phi^{0}\rightarrow\overline{d_{R}}+\widetilde{d}_{L})\nonumber\\
 &\hspace{2cm}+\Gamma(\phi^{0}\rightarrow H_{L}+H_{L}^{\dagger})+\Gamma(\phi^{0}\rightarrow H_{R}+H_{R}^{\dagger}),\nonumber\\
 &\frac{\Gamma(\phi^{0}\rightarrow u_{R}+\overline{\widetilde{u}_{L}})
   -\Gamma(\phi^{0}\rightarrow\overline{u_{R}}+\widetilde{u}_{L})}{\Gamma_{total}(\phi^{0})}
   =\frac{-Im[TrY_{0}Y_{2}^{\dagger}Y_{0}^{T}Y_{1}^{\dagger}]f(x)}
   {8\pi\left(Tr[Y_{1}Y_{1}^{\dagger}+Y_{2}Y_{2}^{\dagger}]+\frac{\mu_{0}^{2}}{M_{\phi^{0}}^{2}}\right)}=\varepsilon,\nonumber\\
 &\frac{\Gamma(\phi^{0}\rightarrow d_{R}+\overline{\widetilde{d}_{L}})
   -\Gamma(\phi^{0}\rightarrow\overline{d_{R}}+\widetilde{d}_{L})}{\Gamma_{total}(\phi^{0})}=\varepsilon,\nonumber\\
 &f(x)=1+2x-2(x+x^{2})ln(1+\frac{1}{x}),\hspace{0.5cm} x=\frac{M_{\phi^{\pm}}^{2}}{M_{\phi^{0}}^{2}}\,.
\end{alignat}
 For $\mu_{0}\sim10^{3}$ GeV, then $\frac{\mu_{0}}{M_{\phi^{0}}}\ll Y_{1,2}\sim10^{-4}$, so the last two decays can indeed be ignored. The second asymmetry has the same result as the first one. In short, the decays of $\phi^{0}$ satisfy two items of Sakharov's three conditions \cite{31}, namely $CP$ violation and out-of-equilibrium.

 In the above decays of $\phi^{0}$, the total baryon number $B+\widetilde{B}$ is undoubtedly conserved, but the $CP$ violation and out-of-equilibrium surely lead to the respective baryon number asymmetries in the $B$ and $\widetilde{B}$ sectors as follow,
\begin{alignat}{1}
 &B_{u_{R}}=(3\times\frac{1}{3})\sum_{i}^{N_{f}}\frac{n_{u_{Ri}}-\overline{n}_{u_{Ri}}}{s}=\kappa\frac{\varepsilon}{g_{*}}\,,\hspace{0.3cm}
  \widetilde{B}_{\widetilde{u}_{L}}=(3\times\frac{1}{3})\sum_{i}^{N_{f}}\frac{n_{\widetilde{u}_{Li}}-\overline{n}_{\widetilde{u}_{Li}}}{s}
  =-\kappa\frac{\varepsilon}{g_{*}}\,,\nonumber\\
 &B_{d_{R}}=(3\times\frac{1}{3})\sum_{i}^{N_{f}}\frac{n_{d_{Ri}}-\overline{n}_{d_{Ri}}}{s}=\kappa\frac{\varepsilon}{g_{*}},\hspace{0.4cm}
  \widetilde{B}_{\widetilde{d}_{L}}=(3\times\frac{1}{3})\sum_{i}^{N_{f}}\frac{n_{\widetilde{d}_{Li}}-\overline{n}_{\widetilde{d}_{Li}}}{s}
  =-\kappa\frac{\varepsilon}{g_{*}},
\end{alignat}
 where $N_{f}$ is fermion generation number, $s$ is entropy density, and $\kappa$ is a dilution factor. For a very weak decay, it is serious departure from thermal equilibrium, so one can approximate $\kappa\approx1$. Note that the asymmetries do not depend on the universe temperature in the comoving volume. Obviously, the $B$ and $\widetilde{B}$ asymmetries are the same size but opposite sign, so the total baryon number asymmetry of the universe is still vanishing. After $\phi^{0}$ decaying and decoupling, the $B$ and $\widetilde{B}$ sectors are out of connection and separated from each other. The $B$ and $\widetilde{B}$ asymmetries will remain in the respective sectors. For $|Y_{0}|\sim 10^{-3}$ and its complex phase $\sim0.1\pi$, the equations of (13) and (14) can correctly give a satisfied baryon asymmetry.

 Things happened next are the sphaleron processes \cite{32}. In the temperature area of $v_{R}<T<M_{\phi^{0}}$, the gauge symmetries are yet unbroken and the mirror symmetry is kept well. In the SM and mirror sector, the baryon and lepton current anomaly and the relevant charge conversion are collected as follows,
\begin{alignat}{1}
 &C_{L}=\frac{N_{f}}{32\pi^{2}}g_{w}^{2}W_{L\mu\nu}^{i}\widetilde{W}_{L}^{i\mu\nu},\hspace{0.3cm}
  C_{R}=\frac{N_{f}}{32\pi^{2}}g_{w}^{2}W_{R\mu\nu}^{i}\widetilde{W}_{R}^{i\mu\nu},\hspace{0.3cm}
  C_{Y}=\frac{N_{f}}{32\pi^{2}}g_{Y}^{2}B_{\mu\nu}\widetilde{B}^{\mu\nu},\nonumber\\
 &J_{\mu}^{B}=\sum_{i}^{N_{f}}(\overline{q_{Li}}\gamma_{\mu}q_{Li}+\overline{u_{Ri}}\gamma_{\mu}u_{Ri}+\overline{d_{Ri}}\gamma_{\mu}d_{Ri}),\nonumber\\
 &J_{\mu}^{\widetilde{B}}=\sum_{i}^{N_{f}}(\overline{\widetilde{q}_{Ri}}\gamma_{\mu}\widetilde{q}_{Ri}
  +\overline{\widetilde{u}_{Li}}\gamma_{\mu}\widetilde{u}_{Li}+\overline{\widetilde{d}_{Li}}\gamma_{\mu}\widetilde{d}_{Li}),\nonumber\\
 &J_{\mu}^{L}=\sum_{i}^{N_{f}}(\overline{l_{Li}}\gamma_{\mu}l_{Li}+\overline{e_{Ri}}\gamma_{\mu}e_{Ri}
  +\overline{\nu_{Ri}}\gamma_{\mu}\nu_{Ri})+(2iS^{\dagger}\partial_{\mu}S+h.c.),\nonumber\\
 &J_{\mu}^{\widetilde{L}}=\sum_{i}^{N_{f}}(\overline{\widetilde{l}_{Ri}}\gamma_{\mu}\widetilde{l}_{Ri}
  +\overline{\widetilde{e}_{Li}}\gamma_{\mu}\widetilde{e}_{Li}+\overline{\widetilde{\nu}_{Li}}\gamma_{\mu}\widetilde{\nu}_{Li})
  +(2i\widetilde{S}^{\dagger}\partial_{\mu}\widetilde{S}+h.c.),\nonumber\\
 &\partial^{\mu}J_{\mu}^{B}=-C_{L}+C_{Y},\hspace{0.5cm}\partial^{\mu}J_{\mu}^{\widetilde{B}}=C_{R}-C_{Y},\nonumber\\
 &\partial^{\mu}J_{\mu}^{L}=-C_{L}+C_{Y},\hspace{0.5cm}\partial^{\mu}J_{\mu}^{\widetilde{L}}=C_{R}-C_{Y},\nonumber\\
 &\Longrightarrow\Delta(B+\widetilde{B})=\Delta(L+\widetilde{L})=0,\hspace{0.5cm}\Delta(B-L)=\Delta(\widetilde{B}-\widetilde{L})=0.
\end{alignat}
 In view of the left-right mirror symmetry, the sphaleron transitions for $SU(2)_{L}$ and $SU(2)_{R}$ are completed in parallel ways. The only difference is the opposite initial asymmetries in the two sectors which are provided by (14). When the universe temperature decreases to $T=v_{R}$, $SU(2)_{R}$ is broken and the mirror sphaleron process is stopped. Therefore, the baryon and lepton asymmetries at $T=v_{R}$ are obtained as follows,
\begin{alignat}{1}
 &(B-L)_{T=v_{R}}=(B-L)_{T=M_{\phi^{0}}}=(B_{u_{R}}+B_{d_{R}})_{T=M_{\phi^{0}}}=2\kappa\frac{\varepsilon}{g_{*}}\,,\nonumber\\
 &(\widetilde{B}-\widetilde{L})_{T=v_{R}}=(\widetilde{B}-\widetilde{L})_{T=M_{\phi^{0}}}
  =(\widetilde{B}_{\widetilde{u}_{L}}+\widetilde{B}_{\widetilde{d}_{L}})_{T=M_{\phi^{0}}}=-2\kappa\frac{\varepsilon}{g_{*}}\,,\nonumber\\
 &B_{T=v_{R}}=c_{sp}(B-L)_{T=v_{R}},\hspace{1.5cm}
  \widetilde{B}_{T=v_{R}}=c_{sp}(\widetilde{B}-\widetilde{L})_{T=v_{R}}=-B_{T=v_{R}},\nonumber\\
 &L_{T=v_{R}}=(c_{sp}-1)(B-L)_{T=v_{R}},\hspace{0.5cm}
  \widetilde{L}_{T=v_{R}}=(c_{sp}-1)(\widetilde{B}-\widetilde{L})_{T=v_{R}}=-L_{T=v_{R}},\nonumber\\
 &c_{sp}=\frac{2N_{f}^{2}+N_{f}}{8N_{f}^{2}+24N_{f}+6}=\frac{7}{50}\:(\mbox{for}\:N_{f}=3),
\end{alignat}
 where $c_{sp}$ is a coefficient of the sphaleron conversion, (16) is in detail derived in appendix A.

 Below the $v_{R}$ scale, $SU(2)_{L}\otimes SU(2)_{R}\otimes U(1)_{Y}\rightarrow SU(2)_{L}\otimes U(1)_{Y'}$, and the left-right mirror symmetry is also lost. Accordingly, the right-handed doublets $\widetilde{q}_{R}$ and $\widetilde{l}_{R}$ are spontaneously decomposed into the uncorrelated states $\widetilde{u}_{R},\widetilde{d}_{R}$ and $\widetilde{\nu}_{R},\widetilde{e}_{R}$, and the right-handed anomaly $C_{R}$ disappears as well. The $B$ sector and the $\widetilde{B}$ one are now connected by virtue of the mixings of $u_{R}$ and $\widetilde{u}_{R}$ as well as $d_{R}$ and $\widetilde{d}_{R}$, but the $L$ sector and the $\widetilde{L}$ one are still separated, see (11). In the temperature area of $v_{L}<T<v_{R}$,  only the sphaleron transition for $SU(2)_{L}$ goes on. The baryon and lepton current anomaly and the relevant charge conversion are now changed as follows,
\begin{alignat}{1}
 &J_{\mu}^{\widetilde{B}}=\sum_{i}^{N_{f}}(\overline{\widetilde{u}_{Ri}}\gamma_{\mu}\widetilde{u}_{Ri}
  +\overline{\widetilde{d}_{Ri}}\gamma_{\mu}\widetilde{d}_{Ri}
  +\overline{\widetilde{u}_{Li}}\gamma_{\mu}\widetilde{u}_{Li}+\overline{\widetilde{d}_{Li}}\gamma_{\mu}\widetilde{d}_{Li}),\nonumber\\
 &J_{\mu}^{\widetilde{L}}=\sum_{i}^{N_{f}}(\overline{\widetilde{\nu}_{Ri}}\gamma_{\mu}\widetilde{\nu}_{Ri}
  +\overline{\widetilde{e}_{Ri}}\gamma_{\mu}\widetilde{e}_{Ri}
  +\overline{\widetilde{e}_{Li}}\gamma_{\mu}\widetilde{e}_{Li}+\overline{\widetilde{\nu}_{Li}}\gamma_{\mu}\widetilde{\nu}_{Li})
  +(2i\widetilde{S}^{\dagger}\partial_{\mu}\widetilde{S}+h.c.),\nonumber\\
 &\partial^{\mu}J_{\mu}^{B}=-C_{L}+C_{Y}+[iv_{\phi}(\overline{u_{R}}Y_{1}\widetilde{u}_{L}
  +\overline{d_{R}}Y_{2}\widetilde{d}_{L})+h.c.],\hspace{0.5cm}\partial^{\mu}J_{\mu}^{L}=-C_{L}+C_{Y},\nonumber\\
 &\partial^{\mu}J_{\mu}^{\widetilde{B}}=-iv_{\phi}(\overline{u_{R}}Y_{1}\widetilde{u}_{L}
  +\overline{d_{R}}Y_{2}\widetilde{d}_{L})+h.c.,\hspace{2.8cm} \partial^{\mu}J_{\mu}^{\widetilde{L}}=0,\nonumber\\
 &\Longrightarrow\Delta(B+\widetilde{B}-L)=\Delta\widetilde{L}=0.
\end{alignat}
 It can be seen from (17) that the mirror baryons have joined the SM sector, so two new separate sectors are now the $B+\widetilde{B}-L$ sector and the $\widetilde{L}$ one. When the energy scale decreases to $v_{L}$, the electroweak breaking occurs and the $SU(2)_{L}$ sphaleron process is stopped. At $T=v_{L}$, the baryon and lepton asymmetries evolve into the below results,
\begin{alignat}{1}
 &(B+\widetilde{B}-L)_{T=v_{L}}=(B+\widetilde{B}-L)_{T=v_{R}}=-L_{T=v_{R}},\hspace{0.4cm}
  \widetilde{L}_{T=v_{L}}=\widetilde{L}_{T=v_{R}}=-L_{T=v_{R}},\nonumber\\
 &(B+\widetilde{B})_{T=v_{L}}=c_{sp}'(B+\widetilde{B}-L)_{T=v_{L}},\hspace{0.5cm}
  L_{T=v_{L}}=(c_{sp}'-1)(B+\widetilde{B}-L)_{T=v_{L}},\nonumber\\
 &(\widetilde{L}_{\widetilde{e}})_{T=v_{L}}=0,\hspace{0.5cm}
  (\widetilde{L}_{\widetilde{\nu}})_{T=v_{L}}=\widetilde{c}_{sp}\widetilde{L}_{T=v_{L}},\hspace{0.5cm}
  (\widetilde{L}_{\widetilde{S}})_{T=v_{L}}=(1-\widetilde{c}_{sp})\widetilde{L}_{T=v_{L}},\nonumber\\
 &c_{sp}'=\frac{10N_{f}^{2}+2N_{f}}{25N_{f}^{2}+45N_{f}+6}=\frac{16}{61}\:(\mbox{for}\:N_{f}=3),\hspace{0.5cm}
  \widetilde{c}_{sp}=\frac{N_{f}}{N_{f}+4}\,,
\end{alignat}
 where $c_{sp}'$ and $\widetilde{c}_{sp}$ are two new coefficients of the sphaleron conversion, a derivation of (18) is in appendix B. Below the $v_{L}$ scale, all kinds of the asymmetries in the $B+\widetilde{B}$ and $\widetilde{L}$ sectors will be kept at all time, but the $L$ sector is not like that because $U(1)_{L}$ is broken at the energy scale $v_{s}\sim100$ GeV. After this, the heavy Majorana neutrino $\nu_{R}$ and the light one $\nu_{L}$ are active in the $L$ sector.

 In the above evolution, the below reactions play key roles,
\begin{alignat}{1}
 &\widetilde{e}^{\,-}+\widetilde{e}^{\,+}\longrightarrow \gamma+\gamma,\hspace{0.2cm}
  \widetilde{e}_{R}^{\,-}+\widetilde{\pi}^{\,+}\longrightarrow \widetilde{\nu}_{R}^{\,0}+\gamma,\hspace{0.2cm}
  \widetilde{e}_{R}^{\,-}+\widetilde{p}^{\,+}\longrightarrow \widetilde{\nu}_{R}^{\,0}+\widetilde{n}^{\,0}
  (\mbox{oscillate into}\:{n}^{0}),\nonumber\\
 &\widetilde{\nu}_{L}+\overline{\widetilde{\nu}_{L}}\longrightarrow \widetilde{S}+\widetilde{S}^{*},\hspace{0.5cm}
  \widetilde{\nu}_{R2,3}\longrightarrow\widetilde{\nu}_{R1}+\gamma,\nonumber\\
 &\nu_{R}+\overline{\nu_{R}}\longrightarrow S_{I}+S_{I},\hspace{0.5cm}
  \nu_{R}\longrightarrow\nu_{L}+e+\overline{e},\hspace{0.5cm} S_{R}\longrightarrow\nu_{R}+\nu_{R}.
\end{alignat}
 Firstly, the heavier mirror baryons can oscillate into the lighter SM baryons due to the mixings between the mirror quarks and the SM quarks. This will eventually lead to the $\widetilde{B}$ asymmetry disappearing. Secondly, the symmetric parts of $\widetilde{e}^{\,-}$ and $\widetilde{e}^{\,+}$ annihilate into photons, while their asymmetric parts convert into the mirror neutrino $\widetilde{\nu}_{R}$ via the reactions with $\widetilde{\pi}^{\,+}$ or $\widetilde{p}^{\,+}$. In particular, the stable mirror atoms can not be formed precisely because of the third reaction in (19). Consequently, all of the mirror baryons, mirror charged leptons and mirror atoms can not survive in the present universe. Thirdly, the symmetric parts of $\widetilde{\nu}$ and $\overline{\widetilde{\nu}}$ annihilate into Goldstone boson pairs of $\widetilde{S}$ and $\widetilde{S}^{*}$, and the heavier $\widetilde{\nu}_{2,3}$ are radiative decay into the lightest $\widetilde{\nu}_{1}$, see Figure 2.
\begin{figure}
 \centering
 \includegraphics[totalheight=6cm]{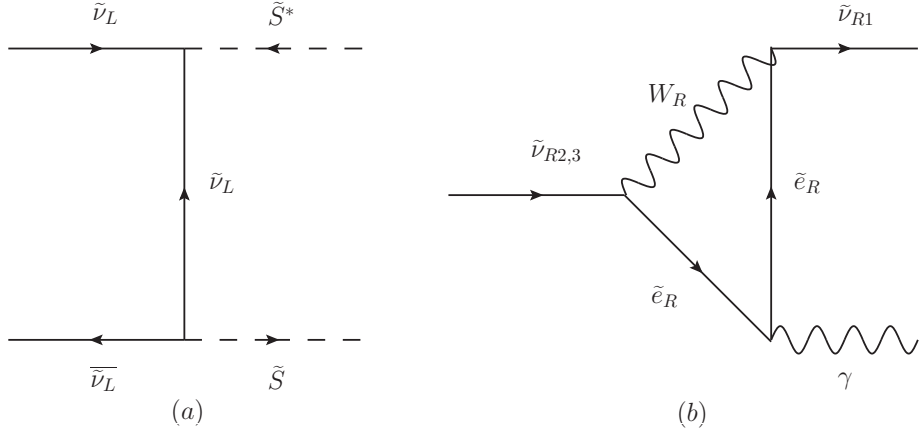}
 \caption{(a) the annihilation way of a pair of $\widetilde{\nu}$, (b) the radiative decay of the heavier $\widetilde{\nu}_{2,3}$ into the lightest $\widetilde{\nu}_{1}$ which is namely the cold dark matter.}
\end{figure}
 Thus only the stable $\widetilde{\nu}_{1}$ survive in the asymmetric parts of $\widetilde{\nu}$, it eventually becomes the cold dark matter in the present universe. Lastly, the two special particles $\nu_{R}$ and $S_{R}$ in the SM sector (of course both of them are not the well-known SM particles) completely annihilate and decay via the third line processes in (19), so they have no relics in the present universe. Finally, the light Majorana neutrino $\nu_{L}$ and the massless Goldstone bosons $S_{I},\widetilde{S}$ are evidently relativistic decoupling, therefore, they become the hot dark matter in the present universe.

 There is no doubt that $\widetilde{\nu}_{1}$ is non-relativistic decoupling. Its freeze temperature is determined by the annihilation cross-section $\sigma(\widetilde{\nu}_{1}+\overline{\widetilde{\nu}_{1}}\longrightarrow \widetilde{S}+\widetilde{S}^{*})$ as follows,
\begin{alignat}{1}
 &\sigma v_{r}=\frac{(Y_{m11}^{*}Y_{m11})^{2}}{16\pi m_{\widetilde{\nu}_{1}}^{2}}
  (1-v^{2})(-5+v^{2}+\frac{3-v^{2}}{v}ln\frac{1+v}{1-v}),\nonumber\\
 &\langle\sigma v_{r}\rangle n_{\widetilde{\nu}_{1}}(T_{\widetilde{\nu}_{1}})
  =H(T_{\widetilde{\nu}_{1}})=\frac{1.66\sqrt{g_{*}}\,T_{\widetilde{\nu}_{1}}^{2}}{M_{pl}},\nonumber\\
 &\Longrightarrow \frac{m_{\widetilde{\nu}_{1}}}{T_{\widetilde{\nu}_{1}}}\approx36+ln\frac{(Y_{m11}Y_{m11}^{*})^{2}}{m_{\widetilde{\nu}_{1}}(GeV)},
\end{alignat}
 where $v_{r}$ is the relative velocity of $\widetilde{\nu}_{1}$ and $\overline{\widetilde{\nu}_{1}}$, and $v$ is the $\widetilde{\nu}_{1}$ velocity in the center-of-mass frame. The heat average can be calculated by $\langle\sigma v_{r}\rangle\approx a+b\langle v^{2}\rangle=a+b\frac{3T_{\widetilde{\nu}_{1}}}{m_{\widetilde{\nu}_{1}}}$ where $T_{\widetilde{\nu}_{1}}$ is the freeze temperature. For $Y_{m}\sim0.1$ and $m_{\widetilde{\nu}_{1}}\sim 1$ GeV, one can estimate $\frac{m_{\widetilde{\nu}_{1}}}{T_{\widetilde{\nu}_{1}}}\approx27$. At this freeze temperature, the relativistic particles include the first generation fermions $u,d,e$, three generations of the light Majorana neutrinos $\nu_{L}$, the Goldstone bosons $S_{I},\widetilde{S}$, and photon. Therefore, $g_{*}$ in (20) should be input by $g_{*}=34.75$. After $\widetilde{\nu}_{1}$ decoupling, the symmetric parts of $\widetilde{\nu}_{1}$ have been exhausted owing to its large annihilation cross-section, only its asymmetric parts survive in the present universe. Finally, it should be noted that $T_{\widetilde{\nu}_{1}}$ is also the decoupling temperature of $\widetilde{S}$, and the $\widetilde{S}$ asymmetry does no change before and after it decoupling.

 To sum up, the current universe matters consist of the photon, baryon, electron, the light Majorana neutrino $\nu_{L}$ and massless Goldstone bosons $S_{I},\widetilde{S}$ which are the hot dark matter, and the mirror Dirac neutrino $\widetilde{\nu}_{1}$ which is the cold dark matter. On account of these matters having a common origin, today their asymmetries and relic abundance have some essential relations as follows,
\begin{alignat}{1}
 &B_{today}=(B+\widetilde{B})_{T=v_{L}},\hspace{0.5cm}
  (\widetilde{L}_{\widetilde{\nu}_{1}})_{today}=(\widetilde{L}_{\widetilde{\nu}})_{T=v_{L}},\hspace{0.5cm}
  (\widetilde{L}_{\widetilde{S}})_{today}=(\widetilde{L}_{\widetilde{S}})_{T=v_{L}},\nonumber\\
 &\eta_{B}=7.04B_{today},\hspace{0.4cm}
  \eta_{\widetilde{\nu}_{1}}=(\frac{n_{\widetilde{\nu}_{1}}-\overline{n}_{\widetilde{\nu}_{1}}}{n_{\gamma}})
  =7.04(\widetilde{L}_{\widetilde{\nu}_{1}})_{today},\hspace{0.4cm}
  \eta_{\widetilde{S}}=7.04(\widetilde{L}_{\widetilde{S}})_{today},\nonumber\\
 \Longrightarrow &\frac{\eta_{\widetilde{\nu}_{1}}}{\eta_{B}}=\frac{\widetilde{c}_{sp}}{c_{sp}'}\approx1.6,\hspace{0.5cm}
  \frac{\eta_{\widetilde{S}}}{\eta_{B}}=\frac{1-\widetilde{c}_{sp}}{c_{sp}'}\approx2.2,\nonumber\\
 &\frac{\Omega_{\widetilde{\nu}_{1}}}{\Omega_{B}}=\frac{m_{\widetilde{\nu}_{1}}\eta_{\widetilde{\nu}_{1}}}{m_{p}\,\eta_{B}}\approx5
  \Longrightarrow\frac{m_{\widetilde{\nu}_{1}}}{m_{p}}\approx 3.1,\hspace{0.5cm}
  \frac{\Omega_{\widetilde{S}}}{\Omega_{\gamma}}=\frac{g_{\widetilde{S}}T_{\widetilde{S}}^{4}}{g_{\gamma}T_{\gamma}^{4}}
  =(\frac{2}{31.75})^{\frac{4}{3}}\approx0.025,
\end{alignat}
 where $7.04$ is a ratio of the entropy density $s$ to the photon number density $n_{\gamma}$. In addition, $\Omega_{S_{I}}$ is smaller than $\Omega_{\widetilde{S}}$ because the decoupling temperature of $S_{I}$ is higher than one of $\widetilde{S}$. However, $\Omega_{\widetilde{S}}$ and $\Omega_{S_{I}}$ are much smaller in comparison with $\Omega_{\nu_{L}}\approx1.7\times10^{-3}$ and $\Omega_{\gamma}\approx5\times10^{-5}$. In view of their same origin, $\eta_{B}$ and $\eta_{\widetilde{\nu}_{1}}$ have a close size, moreover, they are essentially a complementary relationship. At present day, $\eta_{B}$ has been measured, but $\eta_{\widetilde{\nu}_{1}}$ hides itself and eludes all kinds of observations. Obviously, all of the results meet BBN constraints \cite{33}. In a word, (21) are interesting and important predictions of the model. They surely provide a clear guide for the future experimental search.

 Through the above mechanism, the universe has eventually evolved into the final state with both baryon asymmetry and dark matter asymmetry from the initial state with the matter-antimatter symmetry, and it is separated into the visible sector and the dark one. The later numerical results will demonstrate that the model is indeed successful.

\vspace{1cm}
 \noindent\textbf{IV. Numerical Results}

\vspace{0.3cm}
 In the section I present the model numerical results. In the light of the foregoing discussions, the model contains a lot of the new parameters besides the SM ones. In principle the SM parameters have been fixed by the experimental data, but the non-SM parameters have yet large freedoms. In fact, there are not many non-SM parameters involved in the numerical calculations. The gauge sector parameters are the two gauge couplings $g_{w}$ and $g_{Y}$. In view of the relevant relations in (10), I can use the mixing angle $\widetilde{\theta}$ as a substitute for $g_{Y}$, furthermore, $g_{w}$ and $tan\widetilde{\theta}$ are determined by $e$ and $sin\theta_{W}$, which have precisely been measured by the electroweak physics. The scalar sector parameters include the two couplings $\lambda_{H},\lambda_{s}$, the three VEVs $v_{L},v_{R},v_{s}$, and the three mass parameters $M_{\phi^{0}},M_{\phi^{\pm}},\mu_{0}$. Among which, $\lambda_{H}$ and $v_{L}$ are determined by the SM physics. A reasonable value of $\lambda_{s}$ should be around $0.1$. $v_{R}\sim10^{8}$ GeV and $v_{s}\sim100$ GeV are suitable based on the model consistency and experimental limits. $M_{\phi^{0}},M_{\phi^{\pm}}\sim10^{10}$ GeV can satisfy the out-of-equilibrium condition and baryon asymmetry. Finally, $\mu_{0}\sim10^{3}$ GeV leads to $v_{\phi}\sim 0.1$ GeV from (9), this meets multiple requirements of the model. Based on an overall consideration, a set of suitable and typical values of the gauge and scalar parameters are chosen as
\begin{alignat}{1}
 &g_{w}=0.654,\hspace{0.5cm}sin\widetilde{\theta}=0.534,\hspace{0.5cm}\lambda_{H}=0.13,\hspace{0.5cm}\lambda_{s}=0.1,\nonumber\\
 & v_{L}=246\:\mathrm{GeV},\hspace{0.5cm}v_{R}=2\times10^{8}\:\mathrm{GeV},\hspace{0.5cm}v_{s}=100\:\mathrm{GeV},\nonumber\\
 & M_{\phi^{0}}=M_{\phi^{\pm}}=1\times10^{10}\:\mathrm{GeV},\hspace{0.5cm}\mu_{0}=1\times10^{3}\:\mathrm{GeV}.
\end{alignat}
 Now input (22) into the relevant equations in (9) and (10), the gauge and scalar boson masses are straightforward calculated as follows (in GeV as unit),
\begin{alignat}{1}
 &M_{W_{L}}=80.4,\hspace{0.5cm}M_{Z}=91.2,\hspace{0.5cm}M_{W_{R}}=6.5\times10^{7},\hspace{0.5cm}M_{\widetilde{Z}}=7.7\times10^{7},\nonumber\\
 &M_{S_{R}}=44.7,\hspace{0.5cm}M_{H_{L}}=126,\hspace{0.5cm}M_{H_{R}}=1.0\times10^{8}.
\end{alignat}
 $M_{W_{R}}$, $M_{\widetilde{Z}}$ and $M_{H_{R}}$ are dominated by $v_{R}$, while $M_{S_{R}}$ is affected by $v_{s}$. Although the neutral boson $S_{R}$ is lighter than the SM Higgs boson $H_{L}$, it will be difficult to detect it because $S_{R}$ has hardly any interactions with the SM particles.

 The model Yukawa sector contains a great deal of the flavor parameters. However, I can choose such a set of flavor basis that $Y_{e},Y_{m},Y_{1},Y_{2}$ are all real diagonal. $Y_{e}$ is determined by $M_{e}$. Since the flavor structures of $Y_{m},Y_{1},Y_{2}$ are as yet unknown, all of them are taken as constant unit matrices for simplicity. For the same reason, the complex coupling $Y_{0}$ is set as a complex constant unit matrix. Thus, the non-diagonal coupling $Y_{\nu}$ can be given by the lepton mixing matrix $U_{PMNS}$ and three real diagonal parameters, see the below. The Yukawa sector parameters are typically chosen as follows,
\begin{alignat}{1}
 & Y_{1}=Y_{2}=10^{-4}\times I,\hspace{0.5cm} Y_{0}=(3.5\times10^{-3}\times e^{-i\varphi})\times I,\hspace{0.5cm}
   \varphi=0.131\,\pi,\nonumber\\
 & Y_{m}=0.1\times I,\hspace{0.4cm} M_{e}=\mathrm{diag}\left(m_{e},m_{\mu},m_{\tau}\right),\hspace{0.4cm}
   Y_{\nu}=U_{PMNS}\mathrm{diag}\left(y_{1},y_{2},y_{3}\right)U_{PMNS}^{T},\nonumber\\
 & y_{1}=0.203\times10^{-7},\hspace{0.5cm} y_{2}=0.64\times10^{-7},\hspace{0.5cm} y_{3}=1.52\times10^{-7},\nonumber\\
 & sin\theta_{12}=0.558,\hspace{0.5cm} sin\theta_{23}=0.7,\hspace{0.5cm} sin\theta_{13}=0.158,\hspace{0.5cm} \delta^{\,l}=0.
\end{alignat}
 $Y_{1}$ and $Y_{2}$ are limited by the out-of-equilibrium condition (12). The absolute value and complex phase of $Y_{0}$ are in charge of the baryon asymmetry, so their values are obtained by fitting $\eta_{B}$. $Y_{m}\sim0.1$ is very reasonable in the light of (20). $y_{1},y_{2},y_{3}$ are determined by fitting the masses of the cold dark matter $\widetilde{\nu}_{1}$ and the light neutrinos $\nu_{L}$. By use of (11), all masses of the mirror leptons and Majorana neutrinos are obtained as follows,
\begin{alignat}{1}
 & m_{\widetilde{e}_{1}}=415\:\mathrm{GeV},\hspace{0.5cm} m_{\widetilde{e}_{2}}=8.6\times10^{4}\:\mathrm{GeV},\hspace{0.5cm}
   m_{\widetilde{e}_{3}}=1.4\times10^{6}\:\mathrm{GeV},\nonumber\\
 & m_{\widetilde{\nu}_{1}}=2.87\:\mathrm{GeV},\hspace{0.5cm} m_{\widetilde{\nu}_{2}}=9.05\:\mathrm{GeV},\hspace{0.5cm}
   m_{\widetilde{\nu}_{3}}=21.5\:\mathrm{GeV},\nonumber\\
 & m_{\nu_{L1}}=0.049\:\mathrm{eV},\hspace{0.5cm} m_{\nu_{L2}}=0.0088\:\mathrm{eV},\hspace{0.5cm}
   m_{\nu_{L3}}=0.00088\:\mathrm{eV},\nonumber\\
 & m_{\nu_{Ri}}=14.1\:\mathrm{GeV},\hspace{0.5cm}
   \triangle m^{2}_{21}=7.6\times10^{-5}\:\mathrm{eV^{2}},\hspace{0.5cm}
   \triangle m^{2}_{32}=2.37\times10^{-3}\:\mathrm{eV^{2}},
\end{alignat}
 where $\triangle m^{2}_{21}$ and $\triangle m^{2}_{32}$ are the two mass-squared differences of the light Majorana neutrinos. In addition, one can obtain $T_{\widetilde{\nu}_{1}}\approx100$ MeV from (20). By use of (12),(13),(16),(18) and (21), finally the baryon asymmetry and the relic abundance of the cold dark matter $\widetilde{\nu}_{1}$ are calculated as
\ba
 \frac{\Gamma(\phi^{0}\rightarrow u_{R}+\overline{\widetilde{u}_{L}})}{H(T=M_{\phi^{0}})}=0.03,\hspace{0.5cm}
 \eta_{B}=6.15\times10^{-10},\hspace{0.5cm} \frac{\Omega_{\widetilde{\nu}_{1}}}{\Omega_{B}}=5.
\ea
 The above first equation clearly shows that the $\phi^{0}$ decay is indeed out-of-equilibrium. The results in (26) are precisely the current data of the universe observations \cite{34}.

\begin{figure}
 \centering
 \includegraphics[totalheight=8cm]{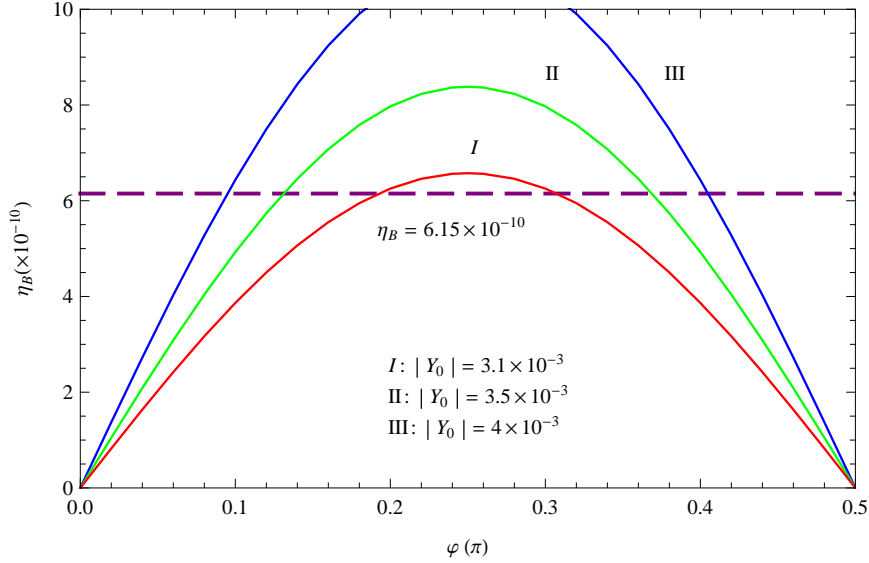}
 \caption{The graphs of the baryon asymmetry subjecting to the phase $\varphi$ for $|Y_{0}|=(3.1\times10^{-3},3.5\times10^{-3},4\times10^{-3})$, while the other parameters are fixed by (22) and (24).}
\end{figure}
 Figure 3 draws $\eta_{B}$ subjecting to $\varphi$ for the three values of $|Y_{0}|=(3.1\times10^{-3},3.5\times10^{-3},4\times10^{-3})$, while the other parameters are fixed by (22) and (24). The left intersection of the curve II and the horizontal baseline of $\eta_{B}=6.15\times 10^{-10}$ exactly corresponds to the values of $|Y_{0}|$ and $\varphi$ in (24). It can be seen from Figure 3 that a reasonable region of $|Y_{0}|$ should be $\sim (3\times10^{-3}-4\times10^{-3})$ for a moderate $\varphi$. In brief, all the numerical results are naturally produced without any fine tuning. They have clearly demonstrated the main ideas of the model.

 In the end, I give a brief discussion about searching the new particles $\widetilde{e},\widetilde{\nu},\nu_{R},S,\widetilde{S}$. On the basis of the model interactions, Figure 4 draws some feasible production processes at the LHC \cite{35}.
\begin{figure}
 \centering
 \includegraphics[totalheight=7cm]{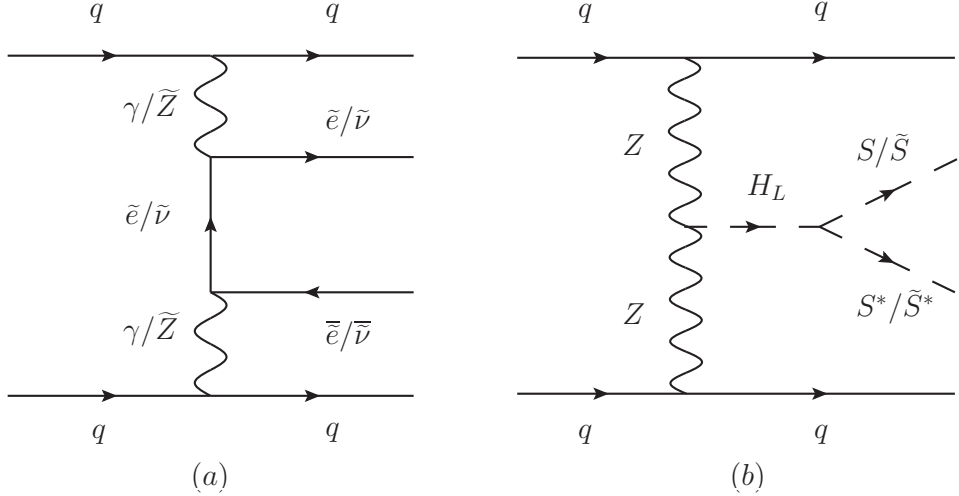}
 \caption{(a) the pair production of $\widetilde{e}$ or $\widetilde{\nu}$ by the proton-proton collisions, (b) the pair production of $S$ or $\widetilde{S}$.}
\end{figure}
 The diagram (a) illustrates the pair production of $\widetilde{e}$ or $\widetilde{\nu}$ by the proton-proton collisions, of course, its cross-section for $\widetilde{\nu}$ is too tiny to be identified. The diagram (b) can produce a pair of $S$ or $\widetilde{S}$, but these cross-sections are tiny due to $S,\widetilde{S}$ having very weak coupling to $H_{L}$. In addition, $H_{L}$ can decay into $\nu_{L}$ and $\nu_{R}$, but it is also very difficult to find $\nu_{R}$ on account of $Y_{\nu}\sim10^{-7}$. The best efficient methods to test the model are of course by the lepton-antilepton collisions at the ILC. The main processes are
\begin{alignat}{1}
 &e^{-}+e^{+}\longrightarrow\gamma\longrightarrow\widetilde{e}^{\,-}+\widetilde{e}^{\,+},\hspace{0.5cm}
  e^{-}+e^{+}\longrightarrow\widetilde{Z}\longrightarrow\widetilde{\nu}+\overline{\widetilde{\nu}},\nonumber\\
 &e^{-}+e^{+}\longrightarrow H_{L}\longrightarrow\nu_{L}+\overline{\nu_{R}},\hspace{0.5cm}
  H_{L}\longrightarrow (S+S^{*})/(\widetilde{S}+\widetilde{S}^{*}).
\end{alignat}
 In particular, the loss of energy in the processes should be regarded as definitive signals of the cold dark matter neutrino $\widetilde{\nu}_{1}$ or the hot dark matter Goldstone bosons $\widetilde{S},S_{I}$. As long as the luminance and running time are enough large, the non-SM particles $\widetilde{e},\nu_{R},S_{R}$ are possible to be discovered. Although all of the searches are large challenging for the future experiments, the model is feasible and promising to be tested in near future.

\vspace{1cm}
 \noindent\textbf{V. Conclusions}

\vspace{0.3cm}
 In the paper, I suggest the left-right mirror symmetric theory to account for the baryogenesis and asymmetric dark matter. The model has the left-right symmetric gauge groups and the global symmetry $U(1)_{B+\widetilde{B}}\otimes U(1)_{L}\otimes U(1)_{\widetilde{L}}$, in addition, the mirror and $Z_{2}$ discrete symmetries. The decays of the super-heavy scalar $\phi^{0}$ are the $CP$ violation and out-of-equilibrium. The $CP$-violating source is the explicit mirror breaking coupling in the Yukawa sector. Through the two steps of the left-right mirror symmetric and asymmetric sphaleron processes, this eventually leads to both the baryon asymmetry and the cold dark matter neutrino asymmetry. The model can not only naturally accommodate the SM and neutrino physics, but also simply and elegantly account for the matter-antimatter asymmetry and dark matter nature, in particular, the both close interrelations are showed. The model gives a number of interesting and important predictions, for instance, the cold dark matter neutrino asymmetry is $1.6$ times the baryon asymmetry, its mass is $3.1$ times the proton mass, and so on. Finally, the model is feasible and promising to be tested in future experiments. Some non-SM particles in the model will possibly be discovered in near future. However, all these efforts will increase our understanding to the mysteries of the universe.

\vspace{1cm}
 \noindent\textbf{Acknowledgments}

\vspace{0.3cm}
 I would like to thank my wife for her large helps. This research is supported by chinese universities scientific fund.

\vspace{1cm}
 \noindent\textbf{Appendix A}

\vspace{0.3cm}
 In a hot and weakly coupled plasma, the asymmetry in the particle and antiparticle number densities is given by its chemical potential. Under the model gauge groups $SU(3)_{C}\otimes SU(2)_{L}\otimes SU(2)_{R}\otimes U(1)_{Y}$, all kinds of the baryon and lepton asymmetries in the SM and mirror sectors are listed as follows,
\begin{alignat}{1}
 &\frac{n_{i}-\overline{n}_{i}}{s}=g_{0}\frac{\mu_{i}}{T}\times
  \left(\begin{array}{c}1\rightarrow\mbox{fermion}\\2\rightarrow\mbox{boson}\end{array}\right),\:
  g_{0}=\frac{45n_{s}}{12\pi^{2}g_{*}}\,,\:n_{s}\:\mbox{is helicity state number},\nonumber\\
 &B_{q_{L}}=2(3\times\frac{1}{3})\sum_{i}^{N_{f}}\frac{n_{q_{Li}}-\overline{n}_{q_{Li}}}{s}=2g_{0}\sum_{i}^{N_{f}}\frac{\mu_{q_{Li}}}{T}\,,\hspace{0.5cm}
  \widetilde{B}_{\widetilde{q}_{R}}=2g_{0}\sum_{i}^{N_{f}}\frac{\mu_{\widetilde{q}_{Ri}}}{T}\,,\nonumber\\
 &B_{u_{R}}=g_{0}\sum_{i}^{N_{f}}\frac{\mu_{u_{Ri}}}{T}\,,\hspace{1cm}
  \widetilde{B}_{\widetilde{u}_{L}}=g_{0}\sum_{i}^{N_{f}}\frac{\mu_{\widetilde{u}_{Li}}}{T}\,,\nonumber\\
 &B_{d_{R}}=g_{0}\sum_{i}^{N_{f}}\frac{\mu_{d_{Ri}}}{T}\,,\hspace{1cm}
  \widetilde{B}_{\widetilde{d}_{L}}=g_{0}\sum_{i}^{N_{f}}\frac{\mu_{\widetilde{d}_{Li}}}{T}\,,\nonumber\\
 &L_{l_{L}}=2g_{0}\sum_{i}^{N_{f}}\frac{\mu_{l_{Li}}}{T}\,,\hspace{1cm}
  \widetilde{L}_{\widetilde{l}_{R}}=2g_{0}\sum_{i}^{N_{f}}\frac{\mu_{\widetilde{l}_{Ri}}}{T}\,,\nonumber\\
 &L_{e_{R}}=g_{0}\sum_{i}^{N_{f}}\frac{\mu_{e_{Ri}}}{T}\,,\hspace{1cm}
  \widetilde{L}_{\widetilde{e}_{L}}=g_{0}\sum_{i}^{N_{f}}\frac{\mu_{\widetilde{e}_{Li}}}{T}\,,\nonumber\\
 &L_{\nu_{R}}=g_{0}\sum_{i}^{N_{f}}\frac{\mu_{\nu_{Ri}}}{T}\,,\hspace{1cm}
  \widetilde{L}_{\widetilde{\nu}_{L}}=g_{0}\sum_{i}^{N_{f}}\frac{\mu_{\widetilde{\nu}_{Li}}}{T}\,,\nonumber\\
 &L_{S}=(-4)g_{0}\frac{\mu_{S}}{T}\,,\hspace{1.3cm}\widetilde{L}_{\widetilde{S}}=(-4)g_{0}\frac{\mu_{\widetilde{S}}}{T}\,,
\end{alignat}
 where $S,\widetilde{S}$ have $(-2)$ lepton numbers. After $\phi^{0}$ decaying and decoupling, the SM sector and the mirror one are separated from each other. The particles in the SM sector are in the thermal reaction equilibrium via the gauge and Yukawa couplings, and the non-perturbative sphaleron processes. There is complete counterparts in the mirror sector in view of the left-right mirror symmetry. Therefore, there are the below relations between the various chemical potentials,
\begin{alignat}{1}
 &\sum_{i}^{N_{f}}(3\mu_{q_{Li}}+\mu_{l_{Li}})=0,\hspace{0.6cm}\sum_{i}^{N_{f}}(3\mu_{\widetilde{q}_{Ri}}+\mu_{\widetilde{l}_{Ri}})=0,\nonumber\\
 &\mu_{q_{Li}}-\mu_{u_{Rj}}+\mu_{H_{L}}=0,\hspace{0.5cm}\mu_{\widetilde{q}_{Ri}}-\mu_{\widetilde{u}_{Lj}}+\mu_{H_{R}}=0,\nonumber\\
 &\mu_{q_{Li}}-\mu_{d_{Rj}}-\mu_{H_{L}}=0,\hspace{0.5cm}\mu_{\widetilde{q}_{Ri}}-\mu_{\widetilde{d}_{Lj}}-\mu_{H_{R}}=0,\nonumber\\
 &\mu_{l_{Li}}-\mu_{e_{Rj}}-\mu_{H_{L}}=0,\hspace{0.6cm}\mu_{\widetilde{l}_{Ri}}-\mu_{\widetilde{e}_{Lj}}-\mu_{H_{R}}=0,\nonumber\\
 &\mu_{l_{Li}}-\mu_{\nu_{Rj}}+\mu_{H_{L}}=0,\hspace{0.6cm}\mu_{\widetilde{l}_{Ri}}-\mu_{\widetilde{\nu}_{Lj}}+\mu_{H_{R}}=0,\nonumber\\
 &\mu_{\nu_{Ri}}+\mu_{\nu_{Rj}}+\mu_{S}=0,\hspace{0.75cm}\mu_{\widetilde{\nu}_{Li}}+\mu_{\widetilde{\nu}_{Lj}}+\mu_{\widetilde{S}}=0,\nonumber\\
 &\sum_{i}^{N_{f}}(\mu_{q_{Li}}-\mu_{l_{Li}}+2\mu_{u_{Ri}}-\mu_{d_{Ri}}-\mu_{e_{Ri}}+\frac{2}{N_{f}}\mu_{H_{L}})=0,\nonumber\\
 &\sum_{i}^{N_{f}}(\mu_{\widetilde{q}_{Ri}}-\mu_{\widetilde{l}_{Ri}}+2\mu_{\widetilde{u}_{Li}}-\mu_{\widetilde{d}_{Li}}
  -\mu_{\widetilde{e}_{Li}}+\frac{2}{N_{f}}\mu_{H_{R}})=0,\nonumber\\
 &\mu_{q_{Li}}\equiv\mu_{q_{L}},\hspace{0.5cm}\mu_{l_{Li}}\equiv\mu_{l_{L}},\hspace{0.5cm}\mu_{u_{Ri}}\equiv\mu_{u_{R}},\hspace{0.3cm}etc.\,,
\end{alignat}
 where the last three lines are hypercharge constrains and generation equilibriums. The chemical potentials can be expressed in terms of $\mu_{l_{L}},\mu_{\widetilde{l}_{R}}$ as follows,
\begin{alignat}{2}
 &\mu_{q_{L}}=-\frac{1}{3}\mu_{l_{L}},&\hspace{0.5cm} &\mu_{\widetilde{q}_{R}}=-\frac{1}{3}\mu_{\widetilde{l}_{R}},\nonumber\\
 &\mu_{u_{R}}=\frac{2N_{f}-1}{6N_{f}+3}\mu_{l_{L}},&\hspace{0.5cm}&\mu_{\widetilde{u}_{L}}=\frac{2N_{f}-1}{6N_{f}+3}\mu_{\widetilde{l}_{R}},\nonumber\\
 &\mu_{d_{R}}=-\frac{6N_{f}+1}{6N_{f}+3}\mu_{l_{L}},&\hspace{0.5cm}&\mu_{\widetilde{d}_{L}}=-\frac{6N_{f}+1}{6N_{f}+3}\mu_{\widetilde{l}_{R}},\nonumber\\
 &\mu_{e_{R}}=\frac{2N_{f}+3}{6N_{f}+3}\mu_{l_{L}},&\hspace{0.5cm}&\mu_{\widetilde{e}_{L}}=\frac{2N_{f}+3}{6N_{f}+3}\mu_{\widetilde{l}_{R}},\nonumber\\
 &\mu_{\nu_{R}}=\frac{10N_{f}+3}{6N_{f}+3}\mu_{l_{L}},&\hspace{0.5cm}&\mu_{\widetilde{\nu}_{L}}=\frac{10N_{f}+3}{6N_{f}+3}\mu_{\widetilde{l}_{R}},\nonumber\\
 &\mu_{H_{L}}=\frac{4N_{f}}{6N_{f}+3}\mu_{l_{L}},&\hspace{0.5cm}&\mu_{H_{R}}=\frac{4N_{f}}{6N_{f}+3}\mu_{\widetilde{l}_{R}},\nonumber\\
 &\mu_{S}=-\frac{20N_{f}+6}{6N_{f}+3}\mu_{l_{L}},&\hspace{0.5cm}&\mu_{\widetilde{S}}=-\frac{20N_{f}+6}{6N_{f}+3}\mu_{\widetilde{l}_{R}}.
\end{alignat}
 Finally this yields the below relations of the baryon and lepton asymmetries,
\begin{alignat}{1}
 &B=B_{q_{L}}+B_{u_{R}}+B_{d_{R}}=\frac{-4}{3}N_{f}g_{0}\frac{\mu_{l_{L}}}{T}\,,\hspace{0.4cm}
  \widetilde{B}=\widetilde{B}_{\widetilde{q}_{R}}+\widetilde{B}_{\widetilde{u}_{L}}+\widetilde{B}_{\widetilde{d}_{L}}
  =\frac{-4}{3}N_{f}g_{0}\frac{\mu_{\widetilde{l}_{R}}}{T}\,,\nonumber\\
 &L=L_{l_{L}}+L_{e_{R}}+L_{\nu_{R}}+L_{S}=\frac{24N_{f}^{2}+92N_{f}+24}{N_{f}(6N_{f}+3)}N_{f}g_{0}\frac{\mu_{l_{L}}}{T}\,,\nonumber\\
 &\widetilde{L}=\widetilde{L}_{\widetilde{l}_{R}}+\widetilde{L}_{\widetilde{e}_{L}}+\widetilde{L}_{\widetilde{\nu}_{L}}+\widetilde{L}_{\widetilde{S}}
  =\frac{24N_{f}^{2}+92N_{f}+24}{N_{f}(6N_{f}+3)}N_{f}g_{0}\frac{\mu_{\widetilde{l}_{R}}}{T}\,,\nonumber\\
 &\Longrightarrow B=c_{sp}(B-L),\hspace{1.4cm}\widetilde{B}=c_{sp}(\widetilde{B}-\widetilde{L}),\nonumber\\
 &\hspace{0.9cm}L=(c_{sp}-1)(B-L),\hspace{0.5cm}\widetilde{L}=(c_{sp}-1)(\widetilde{B}-\widetilde{L}),\nonumber\\
 &\hspace{0.9cm}c_{sp}=\frac{2N_{f}^{2}+N_{f}}{8N_{f}^{2}+24N_{f}+6}=\frac{7}{50}\:(\mbox{for}\:N_{f}=3).
\end{alignat}

\vspace{0.3cm}
 \noindent\textbf{Appendix B}

\vspace{0.3cm}
 After $SU(2)_{L}\otimes SU(2)_{R}\otimes U(1)_{Y}\rightarrow SU(2)_{L}\otimes U(1)_{Y'}$, the left-right mirror symmetry is lost and the right-handed sphaleron process is vanishing. Now the mixing between the $B$ sector and the $\widetilde{B}$ one appears via the $v_{\phi}$ terms, but there is no mixing between the $L$ sector and the $\widetilde{L}$ one. Therefore, the relevant thermal reaction equilibriums are changed as follows,
\begin{alignat}{1}
 &\sum_{i}^{N_{f}}(3\mu_{q_{Li}}+\mu_{l_{Li}})=0,\nonumber\\
 &\mu_{q_{Li}}-\mu_{u_{Rj}}+\mu_{H_{L}}=0,\hspace{0.5cm}\mu_{\widetilde{u}_{Ri}}=\mu_{\widetilde{u}_{Lj}}=\mu_{u_{Rk}},\nonumber\\
 &\mu_{q_{Li}}-\mu_{d_{Rj}}-\mu_{H_{L}}=0,\hspace{0.5cm}\mu_{\widetilde{d}_{Ri}}=\mu_{\widetilde{d}_{Lj}}=\mu_{d_{Rk}},\nonumber\\
 &\mu_{l_{Li}}-\mu_{e_{Rj}}-\mu_{H_{L}}=0,\hspace{0.55cm}\mu_{\widetilde{e}_{Ri}}=\mu_{\widetilde{e}_{Lj}},\nonumber\\
 &\mu_{l_{Li}}-\mu_{\nu_{Rj}}+\mu_{H_{L}}=0,\hspace{0.55cm}\mu_{\widetilde{\nu}_{Ri}}=\mu_{\widetilde{\nu}_{Lj}},\nonumber\\
 &\mu_{\nu_{Ri}}+\mu_{\nu_{Rj}}+\mu_{S}=0,\hspace{0.7cm}\mu_{\widetilde{\nu}_{Li}}+\mu_{\widetilde{\nu}_{Lj}}+\mu_{\widetilde{S}}=0,\nonumber\\
 &\sum_{i}^{N_{f}}\left[\mu_{q_{Li}}-\mu_{l_{Li}}+2(\mu_{u_{Ri}}+\mu_{\widetilde{u}_{Ri}}+\mu_{\widetilde{u}_{Li}})
   -(\mu_{d_{Ri}}+\mu_{\widetilde{d}_{Ri}}+\mu_{\widetilde{d}_{Li}})-\mu_{e_{Ri}}+\frac{2}{N_{f}}\mu_{H_{L}}\right]=0,\nonumber\\
 &\sum_{i}^{N_{f}}\left[0(\mu_{\widetilde{\nu}_{Ri}}+\mu_{\widetilde{\nu}_{Li}})-2(\mu_{\widetilde{e}_{Ri}}+\mu_{\widetilde{e}_{Li}})\right]=0,\nonumber\\
 &\mu_{q_{Li}}\equiv\mu_{q_{L}},\hspace{0.5cm}\mu_{l_{Li}}\equiv\mu_{l_{L}},\hspace{0.5cm}\mu_{u_{Ri}}\equiv\mu_{u_{R}},\hspace{0.3cm}etc.\,.
\end{alignat}
 In terms of $\mu_{l_{L}},\mu_{\widetilde{\nu}_{R}}$, the chemical potentials are expressed as
\begin{alignat}{2}
 &\mu_{q_{L}}=-\frac{1}{3}\mu_{l_{L}},\nonumber\\
 &\mu_{u_{R}}=-\frac{1}{15N_{f}+3}\mu_{l_{L}}=\mu_{\widetilde{u}_{R}}=\mu_{\widetilde{u}_{L}},\nonumber\\
 &\mu_{d_{R}}=-\frac{10N_{f}+1}{15N_{f}+3}\mu_{l_{L}}=\mu_{\widetilde{d}_{R}}=\mu_{\widetilde{d}_{L}},\nonumber\\
 &\mu_{e_{R}}=\frac{10N_{f}+3}{15N_{f}+3}\mu_{l_{L}},\hspace{0.7cm}\mu_{\widetilde{e}_{R}}=\mu_{\widetilde{e}_{L}}=0,\nonumber\\
 &\mu_{\nu_{R}}=\frac{20N_{f}+3}{15N_{f}+3}\mu_{l_{L}},\hspace{0.7cm}
  \mu_{\widetilde{\nu}_{R}}=\mu_{\widetilde{\nu}_{L}}=-\frac{1}{2}\mu_{\widetilde{S}}\,,\nonumber\\
 &\mu_{H_{L}}=\frac{5N_{f}}{15N_{f}+3}\mu_{l_{L}},\nonumber\\
 &\mu_{S}=-\frac{40N_{f}+6}{15N_{f}+3}\mu_{l_{L}}.
\end{alignat}
 Thus, the baryon and lepton asymmetries are given by the relations as follows,
\begin{alignat}{1}
 &B+\widetilde{B}=(B_{q_{L}}+B_{u_{R}}+B_{d_{R}})+(\widetilde{B}_{\widetilde{u}_{R}}+\widetilde{B}_{\widetilde{d}_{R}}
                  +\widetilde{B}_{\widetilde{u}_{L}}+\widetilde{B}_{\widetilde{d}_{L}})
                 =\frac{-8}{3}N_{f}g_{0}\frac{\mu_{l_{L}}}{T}\,,\nonumber\\
 &L=L_{l_{L}}+L_{e_{R}}+L_{\nu_{R}}+L_{S}=\frac{60N_{f}^{2}+172N_{f}+24}{N_{f}(15N_{f}+3)}N_{f}g_{0}\frac{\mu_{l_{L}}}{T}\,,\nonumber\\
 &\widetilde{L}=\widetilde{L}_{\widetilde{\nu}_{R}}+\widetilde{L}_{\widetilde{e}_{R}}+\widetilde{L}_{\widetilde{e}_{L}}
                +\widetilde{L}_{\widetilde{\nu}_{L}}+\widetilde{L}_{\widetilde{S}}
               =\frac{2N_{f}+8}{N_{f}}N_{f}g_{0}\frac{\mu_{\widetilde{\nu}_{R}}}{T}\,,\nonumber\\
 &\Longrightarrow B+\widetilde{B}=c_{sp}'(B+\widetilde{B}-L),\hspace{0.5cm}L=(c_{sp}'-1)(B+\widetilde{B}-L),\nonumber\\
 &\hspace{0.9cm}\widetilde{L}_{\widetilde{e}}=\widetilde{L}_{\widetilde{e}_{R}}+\widetilde{L}_{\widetilde{e}_{L}}=0,\hspace{0.5cm}
 \widetilde{L}_{\widetilde{\nu}}=\widetilde{L}_{\widetilde{\nu}_{R}}+\widetilde{L}_{\widetilde{\nu}_{L}}
  =\widetilde{c}_{sp}\widetilde{L}\,,\hspace{0.5cm}\widetilde{L}_{\widetilde{S}}=(1-\widetilde{c}_{sp})\widetilde{L}\,,\nonumber\\
 &\hspace{0.9cm}c_{sp}'=\frac{10N_{f}^{2}+2N_{f}}{25N_{f}^{2}+45N_{f}+6}=\frac{16}{61}\:(\mbox{for}\:N_{f}=3),\hspace{0.5cm}
  \widetilde{c}_{sp}=\frac{N_{f}}{N_{f}+4}\,.
\end{alignat}


\vspace{1cm}
\begin{thebibliography}{99}
\bibitem{1}
 J. Beringer \emph{et al.} [Particle Data Group], Phys. Rev. D86, 010001 (2012).
\bibitem{2}
 G. Altarelli, M. W. Grunewald, Phys. Reps. 403-404, 189 (2004).
\bibitem{3}
 R. N. Mohapatra, \emph{et al.}, Rep. Prog. Phys. 70, 1757 (2007);
 C. Quigg, Annu. Rev. Nucl. Part. Sci. 59, 505 (2009);
 H. Fritzsch, Int. J. Mod. Phys. A 24, 3354 (2009).
\bibitem{4}
 M. Bartelmann, Rev. Mod. Phys. 82, 331 (2010);
 M. Dine and A. Kusenko, Rev. Mod. Phys. 76, 1 (2004).
\bibitem{5}
 K. S. Babu, arXiv:0910.2948;
 M. Antonelli \emph{et al.}, Phys. Reps. 494, 197 (2010);
 G. Isidori, Y. Nir, and G. Perez, Annu. Rev. Nucl. Part. Sci. 60, 355 (2010).
\bibitem{6}
 A. Hocker and Z. Ligeti, Annu. Rev. Nucl. Part. Sci. 56, 501 (2006);
 L. Camilleri, E. Lisi, and J. F. Wilkerson, Annu. Rev. Nucl. Part. Sci. 58, 343 (2008);
 R. Brugnera, Int. J. Mod. Phys. A 26, 4901 (2011).
\bibitem{7}
 T. Schwetz, M. Tortola and J. W F Valle, New J. Phys 10, 113011 (2008);
 Y. Fukuda \emph{et al.} [Super-Kamiokande Collaboration], Phys. Rev. Lett. 81, 1562 (1998); Phys. Rev. Lett. 85, 3999 (2000);
 M. Apollonio \emph{et al.} [CHOOZ Collaboration], Phys. Lett. B 466, 415 (1999); Eur. Phys. J. C 27, 331 (2003);
 K. Eguchi \emph{et al.} [KamLAND Collaboration], Phys. Rev. Lett. 90, 021802 (2003);
 Q. R. Ahmad \emph{et al.} [SNO Collaboration], Phys. Rev. Lett. 89, 011301 (2002).
\bibitem{8}
 F. T. Avignone III, S. R. Elliott, J. Engel, Rev. Mod. Phys. 80, 481 (2008);
 J. D. Vergados, H. Ejiri and F. Simkovic, Rep. Prog. Phys. 75, 106301 (2012).
\bibitem{9}
 G. C. Branco, R. Gonzalez Felipe, F. R. Joaquim, Rev. Mod. Phys. 84, 515 (2012);
 H. Nunokawa, S. Parke and J. Valle, Prog. Part. Nucl. Phys. 60, 338 (2008).
\bibitem{10}
 L. Canetti, M. Drewes and M. Shaposhnikov, New J. Phys. 14, 095012 (2012);
 J. M. Cline, arXiv:hep-ph/0609145v3.
\bibitem{11}
 G. Bertone, D. Hooper and J. Silk, Phys. Rep. 405, 279 (2005);
 D. Hooper, arXiv:0901.4090;
 E. Aprile and S. Profumo, New J. Phys. 11, 105002 (2009).
\bibitem{12}
 Z. Z. Xing, Nucl. Phys. B (Proc. Suppl.) 203¨C204, 82 (2010);
 S. T. Petcov, Int. J. Mod. Phys. A 25, 4325 (2010);
 P. Langacker, Annu. Rev. Nucl. Part. Sci. 62, 215 (2012);
 Y. Farzan, Int. J. Mod. Phys. A 26, 2461 (2011);
 R. W. Schnee, arXiv:1101.5205.
\bibitem{13}
 G. Altarelli, arXiv:hep-ph/0610164;
 C. H. Albright and M. C. Chen, arXiv:hep-ph/0608137;
 M. C. Chen and K. T. Mahanthappa, Int. J. Mod. Phys. A 18, 5819 (2003);
 W. M. Yang, Phys. Rev. D 87, 095003 (2013);
 W. M. Yang and H. H. Liu, Nucl. Phys. B 820, 364 (2009);
 W. M. Yang and Z. G. Wang, Nucl. Phys. B 707, 87 (2005).
\bibitem{14}
 James M. Cline, arXiv:hep-ph/0609145;
 D. E. Morrissey and M. J. Ramsey-Musolf, New J. Phys. 14, 125003 (2012).
\bibitem{15}
 M. Fukugita, T. Yanagida, Phys. Lett. B 174, 45 (1986);
 S. Davidson, E. Nardi and Y. Nir, Phys. Reps. 466, 105 (2008).
\bibitem{16}
 K. Dick, M. Lindner, M. Ratz and D. Wright, Phys. Rev. Lett. 84, 4039 (2000).
\bibitem{17}
 T. Bringmann, New J. Phys. 11, 105027 (2009);
 C. P. Burgess, M. Pospelov and T. ter Veldhuis, Nucl.Phys.B619, 709 (2001).
\bibitem{18}
 L. Canetti, M. Drewes and M. Shaposhnikov, Phys. Rev, Lett 110, 061801 (2003);
 A. Boyarsky, O. Ruchayskiy, and M. Shaposhnikov, Annu. Rev. Nucl. Part. Sci. 59, 191 (2009);
 A. Kusenko, Phys. Reps. 481, 1 (2009).
\bibitem{19}
 G. Jungman, M. Kamionkowski, and K. Griest, Phys. Reps. 267, 195 (1996).
\bibitem{20}
 G. Bertone, Particle Dark Matter (Cambridge University Press, 2010).
\bibitem{21}
 K. M. Zurek, Phys. Reps. 537, 91 (2014);
 K. Petraki, R.R. Volkas, Int. J. Mod. Phys. A 28, 1330028 (2013).
\bibitem{22}
 A. Davidson and K. C. Wali, Phys. Rev. Lett. 59, 393 (1987);
 K. S. Babu and R. N. Mohapatra, Phys. Rev. Lett. 62, 1079 (1989).
\bibitem{23}
 see reviews in Particle Data Group.
\bibitem{24}
 V. A. Kuzmin, V. A. Rubakov, M. A. Shaposhnikov, Phys. Lett. B 155, 36 (1985).
\bibitem{25}
 M. Carena and H. E. Haber, Prog. Part. Nucl. Phys. 50, 63 (2003);
 A. Djouadi, arXiv:1203.4199.
\bibitem{26}
 ATLAS Collab., Phys. Lett. B710, 49 (2012);
 CMS Collab., Phys. Lett. B710, 26 (2012).
\bibitem{27}
 M. Gell-Mann, P. Ramond, R. Slansky, in \emph{Supergravity}, eds. P. van Niewenhuizen and D. Z. Freeman
 (North-Holland, Amsterdam, 1979);
 T. Yanagida, in \emph{Proc. of the Workshop on Unified Theory and Baryon Number in the Universe},
 eds. O. Sawada and A. Sugamoto (Tsukuba, Japan, 1979);
 R. N. Mohapatra, G. Senjanovic, Phys. Rev. Lett. 44, 912 (1980).
\bibitem{28}
 M. Kobayashi and T. Maskawa, Prog. Theor. Phys. 49, 652 (1973).
\bibitem{29}
 B. M. Pontecorvo, Sov. Phys. JETP 6, 429 (1958);
 Z. Maki, M. Nakagawa and S. Sakata, Prog. Theor. Phys. 28, 870 (1962).
\bibitem{30}
 R. Allahverdi, R. Brandenberger, F. Cyr-Racine, and A. Mazumdar, Annu. Rev. Nucl. Part. Sci. 60, 27 (2010).
\bibitem{31}
 A. D. Sakharov, Pisma Zh. Eksp. Teor. Fiz. 5, 32 (1967) [JETP Lett. 5, 24 (1967 SOPUA,34,392-393.1991 UFNAA,161,61-64.1991)].
\bibitem{32}
 W. Buchmuller, R. D. Peccei and T. Yanagida, Annu. Rev. Nucl. Part. Sci. 55, 311 (2005);
 D. S. Gorbunov and V. A. Rubakov, Introduction to The Theory of The Early Universe: Hot Big Bang Theory (World Scientific Publishing Co. Pte. Ltd, 2011).
\bibitem{33}
 see reviews in Particle Data Group;
 M. Pospelov and J. Pradler, Annu. Rev. Nucl. Part. Sci. 60, 539 (2010).
\bibitem{34}
 E. Komatsu \emph{et al} [WMAP Collaboration], Astrophys. J. Suppl. 180, 330 (2009);
 D. N. Spergel \emph{et al.}, Ap. J. Supp. 148, 175 (2003);
 H. W. Hu, S. Dodelson, Ann. Rev. Astron. Astrophys. 40, 171 (2002).
\bibitem{35}
 David E. Morrissey, Tilman Plehn c, Tim M.P. Tait, Phys. Reps. 515, 1 (2012).
\end{thebibliography}
\end{document}